\documentclass[12pt]{article}
\usepackage{amsmath}
\usepackage{graphicx}
\usepackage{epsfig}
\usepackage{amsfonts}
\usepackage{slashed}

\newcommand{\br}{\hskip .25cm/\hskip -.25cm}

\newcommand{\ol}{\overline}

\newcommand{\lapproxeq}{\lower .7ex\hbox{$\;\stackrel{\textstyle
<}{\sim}\;$}}
\newcommand{\gapproxeq}{\lower .7ex\hbox{$\;\stackrel{\textstyle
>}{\sim}\;$}}
\newcommand{\stackdown}[2]{\lower 1.4ex\hbox{$\;\stackrel{\textstyle{#1}}
{\scriptstyle{#2}}\;$}}
\newcommand{\be}{\begin{equation}}
\newcommand{\ee}{\end{equation}}
\newcommand{\beq}{\begin{equation}}
\newcommand{\eeq}{\end{equation}}
\newcommand{\bea}{\begin{eqnarray}}
\newcommand{\eea}{\end{eqnarray}}

\makeatletter

\def\slash{\@ifnextchar[{\fmsl@sh}{\fmsl@sh[0mu]}}
\def\fmsl@sh[#1]#2{%
  \mathchoice
    {\@fmsl@sh\displaystyle{#1}{#2}}%
    {\@fmsl@sh\textstyle{#1}{#2}}%
    {\@fmsl@sh\scriptstyle{#1}{#2}}%
    {\@fmsl@sh\scriptscriptstyle{#1}{#2}}}
\def\@fmsl@sh#1#2#3{\m@th\ooalign{$\hfil#1\mkern#2/\hfil$\crcr$#1#3$}}
\makeatother

\def\beq{\begin{equation}}
\def\eeq{\end{equation}}

\catcode`\@=11 

\def\lsim{\mathrel{\mathpalette\@versim<}}
\def\gsim{\mathrel{\mathpalette\@versim>}}
\def\@versim#1#2{\vcenter{\offinterlineskip
    \ialign{$\m@th#1\hfil##\hfil$\crcr#2\crcr\sim\crcr } }}

\def\t1{{\tilde 1}}

\def\slash#1{#1\hskip-6pt/\hskip6pt}

\def\to{\rightarrow}

\begin{document}

\begin{titlepage}

\begin{flushright}
CERN-TH-2017-123\\
KCL-PH-TH/2017-29 \\
ACT-05-17, MI-TH-1757
\end{flushright}
\vspace{1.5cm}
\begin{centering}

{\large  
\textbf{ A Novel Foamy Origin for Singlet Fermion Masses}}
\vspace{1cm}

{\bf John Ellis$^{a,b}$}, {\bf Nick~E.~Mavromatos$^{a}$} and {\bf Dimitri V. Nanopoulos}$^{c}$

\vspace{0.4cm}

$^a$ {Theoretical Particle Physics and Cosmology Group, Department of Physics, King's College London,
Strand WC2R 2LS, London, U.K.}

$^b$ {Theoretical Physics Department, CERN, CH-1211 Geneva 23, Switzerland.}

$^{c}$ George P. and Cynthia W. Mitchell Institute for Fundamental
Physics and Astronomy, 
Texas A \& M University, College Station, TX 77843, USA; \\
Astroparticle Physics Group, Houston Advanced Research Center (HARC), Mitchell Campus, Woodlands, TX 77381, USA; \\
Division of Natural Sciences, Academy of Athens, Athens 106 79, Greece
\vspace{1.5cm}

{\bf Abstract}

\end{centering}

\vspace{0.8cm}

We show how masses for singlet fermions can be generated
by interactions with a D-particle model of space-time foam inspired by brane theory.
It has been shown previously by one of the authors (N.E.M.) such interactions may
generate generate dynamically small masses for charged fermions via the recoils of D-particle
defects interacting with photons. In this work we consider the direct interactions of D-particle with 
uncharged singlet fermions such as right-handed neutrinos. Quantum fluctuations of the lattice of 
D-particles have massless vector (spin-one) excitations that are analogues of phonons. These
mediate forces with the singlet fermions, generating large dynamical masses that may be
communicated to light neutrinos via the seesaw mechanism.

\vfill
\begin{flushleft}
June 2017 
\end{flushleft}

\end{titlepage}

\setcounter{equation}{0}

\section{Introduction and motivation}

It is widely accepted that neutral singlet particles with no symmetries to protect them,
such as right-handed neutrinos, will in general have very large masses that might
approach the Planck mass at which quantum-gravitational effects become important.
It has long been argued that quantum-mechanical effects generate a foamy structure
of space-time on small scales~\cite{wheeler}, interactions with which might have observable implications
for the propagation of particles~\cite{nature}. Since foam constitutes a fluctuating background, it is not
obvious whether the conventional rules of quantum field theory apply~\cite{notQFT}, and one possibility
is that interactions with the foam might induce Lorentz-~\cite{kostel} and/or CPT-violating~\cite{sarkarbeny,LVbounds} dispersion relations for 
photons~\cite{nature} and gravitons~\cite{grav}. 

Here we investigate the more conservative possibility
that interactions with space-time foam might contribute to the large masses expected for
neutral singlet particles.

We have long advocated a toy model of space-time foam, motivated by string 
theory~\cite{dfoam,westmuckett,sakharov,emnnewuncert,li}, in which the foamy space-time
structures are provided by stringy defects, D-particles, that interact with matter particles
described by string excitations on branes. These fluctuating D-foam defects break Poincar\'e 
invariance, and their recoil during the interaction with propagating string states breaks Lorentz
invariance. The non-trivial transfer of momentum during the interactions of matter strings 
with D-foam defects is mediated by the emission of non-local intermediate string states that
do not admit a local effective action description, leading to a violation of Lorentz invariance
that is subject to probes using astrophysical sources. The ultraviolet completion of this type
of foam model is provided by string/brane theory itself~\cite{string}.

It was argued in~\cite{mavrolorentz} that another observable implication of D-foam
would be the dynamical generation of small non-perturbative masses for charged fermions. 
This was the result of Lorentz-violating higher-derivative terms in the Maxwell action induced 
by the interaction of photons with the D-particles, which in turn are communicated to the charged
sector via the coupling with the electromagnetic field~\footnote{Due to charge conservation, a
direct coupling of a charged fermion to the D-foam excitations is forbidden by the 
electromagnetic U(1) gauge symmetry~\cite{sakharov}.}. The effective field theory model
upon which this effect is based, onto which the D-foam/fermion/photon effective action is mapped
at low energies, ignoring the non-local stringy structures of the D-particles, has been studied
in a different context in~\cite{alexandre}.

It was argued in~\cite{mavrolorentz} that the quantum fluctuations of the D-particle defects 
in target space lead to a novel {correspondence principle}, through which an antisymmetric
tensor background field in phase space, representing the recoil velocity of the defect during its 
interaction with matter, is mapped~\cite{szabo} into a spatial derivative operator along the 
direction of the recoil. In this way, the resulting Finsler-type~\cite{finsler} Born-Infeld (FBI) Lagrangian
that describes the low-energy dynamics of open strings on a brane world in interaction with the D-particles
may be transformed into an effective Lagrangian with Lorentz-violating higher-order spatial-derivative terms,
reproducing the minimal Lorentz-violating modification of QED  considered  in~\cite{alexandre}. 
It was further argued that this model provide a novel way of generating charged-particle
masses dynamically. We should stress that this model is not of Horava-Lifshitz type~\cite{horava}, 
in the sense that there is no anisotropic scaling between time and space coordinates. The
Lorentz violation is manifested through higher-order spatial-derivative terms that respect rotational symmetry
 in three-space, but are suppressed by an effective mass scale. The presence of this scale and the
 Lorentz-violating terms generate dynamically small masses for charged fermions, for arbitrarily weak gauge fields. 

Our link~\cite{mavrolorentz} of such a Lagrangian with D-foam, and ultimately with more general
models of quantum-gravitational foam, constitutes an explicit realization of the effects of a foam 
medium in slowing down some particles via mass generation. In our case, the mass scale that
suppresses the Lorentz-violating terms is expressed in terms of the string mass scale and 
fundamental parameters of the foam such as the variance of its quantum fluctuations. There are 
issues associated with a quantum-ordering ambiguity in our construction, which was argued 
in~\cite{mavrolorentz} to be resolved by appealing to the important r\^ole of the dynamical 
mass generation in curing infrared instabilities in the model. In this way, a selection of the
physically relevant class of quantum orderings could be made, resulting~\cite{mavrolorentz} in  
the above-mentioned mapping of the quantum-ordered FBI effective action onto the QED 
action of~\cite{alexandre} at low energies. Any further ordering ambiguities within this class of 
models are absorbed into the quantum fluctuation parameters of the D-foam, which are regarded
as phenomenological at this stage.

In this paper we extend this model of D-foam and its interactions with stringy matter to
show how they may generate large masses for neutral singlet fermions such as right-handed
neutrinos via spin-one excitations of the space-time foam that are analogous to phonons
in condensed-matter physics. If the singlet fermions are Dirac, we find that the foam
interactions can be described by the minimal Lorentz-violating 
effective action of \cite{alexandre}, whereas if the singlet fermions are Majorana, 
as might be the case for right-handed neutrinos, they can be described by the Lorentz-violating low-energy 
effective action of \cite{leite}.

The structure of the article is as follows: in the next Section \ref{sec:dfoam} we review the 
mapping of the low-energy dynamics of D-foam onto an appropriately quantum-ordered 
Finsler-Born-Infeld action, which yields the minimal Lorentz-violating QED-type model 
of~ \cite{alexandre,leite}, following the procedure  of  \cite{mavrolorentz}. The novelty in the 
dynamical generation of masses for neutral singlet fields, as compared to the charged fermion case
considered in~\cite{mavrolorentz}, is that the quantum fluctuations of the D-particles themselves 
(which are represented by vector (spin-one) quantum fields) couple directly to the singlet
fields via minimal couplings that resemble the corresponding QED fermion-photon coupling.
However, this vector field is not a photon, nor is it a conventional gauge field if the singlet fields
are Majorana fermions, as is often assumed in quantum field theory models. We discuss in
Section \ref{sec:3} how this coupling generates a dynamical mass for the singlet fermions in 
this model, expressing the obtained masses in terms of parameters of the D-foam model, 
such as the string scale, the density of D-foam excitations and the variance of the quantum 
fluctuations. Some phenomenological issues related to the sizes of the proposed masses,
and the possibility of their enhancement in some multibrane-world scenarios, are discussed
briefly in Section \ref{sec:4}. Our conclusions and an outlook are presented in Section \ref{sec:concl}.

\section{A Model for D-foam Interactions with Singlet Fermions \label{sec:dfoam}}

In this section we discuss an effective field theory model for the Lorentz-violating vector interactions of a
singlet fermion with quantum-fluctuating D-particles. We demonstrate that Lorentz-violating Lagrangians 
of the form discussed in~\cite{alexandre} may arise as (parts of) the low-energy, weak-field limit of an 
effective Born-Infeld Lagrangian describing the propagation of vector bosons ${\mathcal A}_\mu$
that describe the quantum fluctuations of the D-particle foam. We assume for simplicity a uniform lattice of
D-particle defects that puncture a three-dimensional brane world (D3-brane), through which the singlet
fermions propagate. The vector $\mathcal A_\mu$ is the low-energy mode of open strings stretched between the 
D-particle and the D3-brane world and, like the phonons in condensed-matter physics,
it plays the r\^ole of a Goldstone boson. However, it is a vector boson, as a result of the spontaneous 
breaking of translational invariance by fluctuating D-particle defects in the D-foam vacuum.

As already mentioned, such models have been termed D-particle 
foam~\cite{dfoam,westmuckett,emnnewuncert,li}, and have a variety of applications, ranging from providing
microscopic string-inspired models of Lorentz-violating vacuum refractive indices~\cite{emnnewuncert,li,review}
to enhancing the abundance of thermal dark matter relics, with interesting implications for astroparticle 
phenomenology~\cite{vergou}.

\subsection{Features of the D-Foam Model}

\begin{figure}[ht]
\centering
\includegraphics[width=6.5cm]{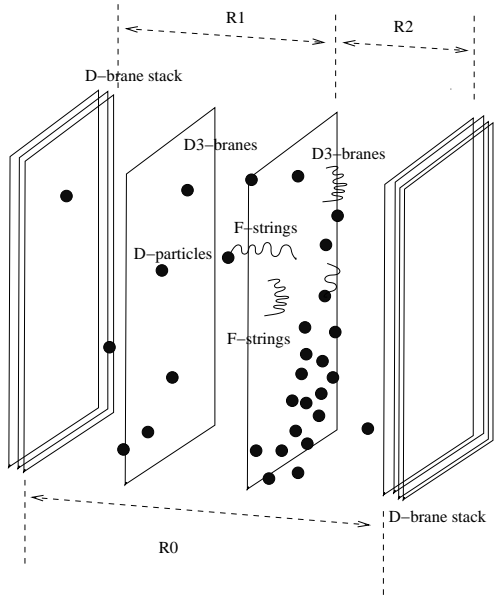}
\caption{\it Schematic
representation of the D-particle model of space-time foam proposed in~\cite{westmuckett}. There are two stacks of D8-branes,
each attached to an orientifold plane, whose special reflective properties enable them to provide
a natural compactification of the bulk dimension, and the bulk is punctured by D0-branes (D-particles).
An isolated D-particle cannot exist in the absence of a D-brane, because of gauge flux conservation. 
Standard Model matter is represented by pen strings on the brane world,  
whereas gravitons and singlet fermions can propagate in the bulk. The latter can interact directly with the 
D-particles in a topologically non-trivial way, involving cutting and splitting the open string representing the singlet fermion.}%
\label{fig:dfoam}%
\end{figure}

The basic idea of the D-foam model is shown in Fig.~\ref{fig:dfoam}.
A (possibly compactified from higher dimensions) three-brane world is moving in a higher-dimensional bulk space,
that is punctured by D-particles. Depending on the string theory considered, the latter can be either point-like 
D0-branes (as in Type IIA strings)~\cite{dfoam,westmuckett,emnnewuncert} or D3-branes wrapped around 
appropriate three cycles (as in Type IIB strings~\cite{li}). As the brane world and the D-particles move in the bulk, 
they may encounter each other and an observer on the D3-brane sees the D-particle defects as flashing 
on and off. In this scenarios, ordinary matter and radiation are represented by open strings with their ends 
attached on the D3-brane.

Gravitational degrees of freedom and singlet fermions are allowed to propagate 
in the bulk~\cite{arkani}, in contrast to SM fields, which are confined to the brane. 
However, when the bulk space is compactified, e.g., through periodic boundary conditions imposed on 
various brane worlds that populate the bulk, see Fig.~\ref{fig:dfoam}, these fermions acquire 
Kaluza-Kilein masses that are suppressed by the size of the extra bulk dimension. 
This mechanism can generate Majorana masses, e.g., for right-handed neutrinos~\cite{arkani,rizos} in brane world scenarios. 
Here we assume that such a mechanism either is not in operation, because our singlet fermions are also
confined on the brane world, like the Standard Model excitations, or that the size of the extra dimensions 
are sufficiently large that the Kaluza-Klein generated masses are negligible. In this case, 
the dominant source of mass for the singlet fermions may be their interaction with the foam. 

Indeed, there are non-trivial interactions of D-particles with open strings, provided there is no electric flux along 
the open string excitations, i.e., provided that the string states are electrically neutral~\footnote{In the
case of Type IIB strings, where the D-particles are not point-like, there may be interactions between 
electrically-charged excitations and the D-particles, but the foam effects on such charged particles are 
suppressed~\cite{li} compared to those on neutral particles. Therefore, in our discussion below we do
not differentiate between these two cases, except when we discuss some phenomenological issues in 
Section \ref{sec:4}.}. This is because the electrically-neutral D-particles can `cut' an open string, 
leading to the emission of intermediate strings stretched between the D-particle and the D3 
brane~\cite{emnnewuncert,review}. If the open string state carries electric flux, such a cutting procedure
is not allowed, due to charge conservation. As a result of the local SU(2) gauge symmetry of 
lepton doublets, this also precludes light (active) neutrinos from interacting directly with the foam.
However, singlet fermions such as right-handed neutrinos may interact directly with the D-particles
as can photons and gravitons. 

As discussed in \cite{review}, the first-quantization picture for the interaction of an open-string
state such as a photon with a D-particle is provided by a world-sheet $\sigma$-model with the following deformation:
\begin{equation}
\mathcal{V}_{\rm{recoil~velocity~part}}^{\rm impulse}=\frac{1}{2\pi\alpha '}
\sum_{i=1}^{D}\int_{\partial D}d\tau\,u_{i}%
X^{0}\Theta\left(  X^{0}\right)  \partial_{n}X^{i} \, , \label{fullrec}%
\end{equation}
where $M_s$ is the string (mass) scale, $g_s$ is the string coupling, and $u_i$ is the
recoil velocity of the D-particle. The latter is assumed to be heavy, and $u^i$ is the spatial part of the
four-velocity of the D-particle, which is well approximated by the ordinary velocity, to leading order
for non-relativistic, slow-moving, heavy D-particles. The  limit $D$ in the sum in (\ref{fullrec})
denotes the appropriate number of spatial target-space dimensions, which is $D = 3$ for a 
recoiling D-particle confined on a D3 brane, as is our case here.
The operator $\Theta_\varepsilon (X^0) = -i \int_{-\infty}^\infty \frac{d\omega}{\omega + i\varepsilon}$:
$\varepsilon \rightarrow 0^+ $, is a regularized Heaviside world-sheet operator.

The recoil operators satisfy~\cite{recoil,szabo} a specific type of conformal algebra, termed a logarithmic conformal algebra~\cite{lcft}. 
This algebra is the limiting case of world-sheet 
algebras that can still be classified by conformal blocks. The impulse operator
$\Theta(X^0)$ is regularized so that the logarithmic conformal field theory algebra is respected~\footnote{This 
can be done by using the world-sheet scale, $\varepsilon^{-2} \equiv {\rm ln}\left(L/a\right)^2$, 
with $a$ an ultraviolet scale and $L$ the world-sheet area, as a regulator~\cite{recoil,szabo}: 
$\Theta_\varepsilon (X^0) = -i\,\int_{-\infty}^\infty \frac{d\omega}{\omega- i\varepsilon} e^{i\omega X^0}$,
taking the quantity $\varepsilon \to 0^+$ at the end of the calculations.}.
The conformal algebra is consistent with momentum conservation during recoil~\cite{recoil,szabo}, which 
yields the following expression for the recoil velocity $u_i$ in terms of momentum transfer during the scattering:
\begin{equation}
u_i = g_s\frac{p_1 - p_2}{M_s}~,
\label{recvel}
\end{equation}
with ${M_s}/{g_s}$ being the D-particle ``mass'' and $\Delta p \equiv p_1 - p_2$ the 
momentum transferred to the string state by its scattering  with the D-particle.

We note next that one can write the boundary recoil/capture operator (\ref{fullrec}) as a total derivative 
over the bulk of the world-sheet, by means of the two-dimensional version of Stokes theorem. 
Omitting from now on the explicit summation over repeated $i$-indices, which is understood to be 
over the spatial indices of the D3-brane world, we then write
\begin{eqnarray}\label{stokes}
&& \mathcal{V}_{\rm{recoil~velocity~part}}^{\rm impulse}=\frac{1}{2\pi\alpha '}
\int_{D}d^{2}z\,\epsilon_{\alpha\beta} \partial^\beta
\left(  \left[  u_{i}X^{0}\right]  \Theta_\varepsilon \left(  X^{0}\right)  \partial^{\alpha}X^{i}\right) = \nonumber \\
&& \frac{1}{4\pi\alpha '}\int_{D}d^{2}z\, (2u_{i})\,\epsilon_{\alpha\beta}
 \partial^{\beta
}X^{0} \Bigg[\Theta_\varepsilon \left(X^{0}\right) + X^0 \delta_\varepsilon \left(  X^{0}\right) \Bigg] \partial
^{\alpha}X^{i} \, ,
\end{eqnarray}
where $\delta_\varepsilon (X^0)$ is an $\varepsilon$-regularized $\delta$-function.

We consider relatively large times after the after the moment of impulse, $X^0 = 0$, at which the initial
open-string state splits into intermediate open strings as a result of the topologically non-trivial 
interactions with the D-particle. For the phenomenological purposes of this work, 
the expression (\ref{stokes}) is equivalent to a deformation describing an open string propagating 
in an antisymmetric  $B_{\mu\nu}$-background ($B$-field) corresponding to a constant 
external  ``electric'' field in target-space:
\begin{equation}\label{Bfoam}
T^{-1} \, B_{i0} = - T^{-1}\, B_{0i} =  u_i = \frac{g_s \,\Delta p_i}{M_s}~, \qquad T=\frac{1}{2\pi \alpha '}~,
\end{equation}
where $T$ denotes the (open) string tension, $0$ is a temporal index, and $i$ is a spatial index. 
The reader should notice here the phase-space dependence of the background field, 
which resembles an `electric' field background but in momentum space~\cite{review}, 
and is therefore of Finsler type~\cite{finsler}.

In the above analysis we have ignored a possible angular momentum operator, 
which may also arise as a result of the non-trivial scattering of a photon on a D-particle defect. 
At the $\sigma$-model level, the latter is also described by a logarithmic conformal algebra deformation
that, for the three-brane case to which we restrict our attention here, takes the form:
\begin{eqnarray}
&& V^{\rm impulse}_{\rm ang~mom~D-part} \; = \; T^{-1} \, \int_{\partial \Sigma} u^i \epsilon_{ijk} X^j \Theta_\varepsilon (X^0) \, \partial_n X^k \nonumber \\ &=& T^{-1}\,\int_{\rm \Sigma} \varepsilon^{\alpha\beta} \left( u^i \epsilon_{ijk} \Theta_\varepsilon (X^0) \, \partial_\alpha X^j \partial_\beta X^k
+ u^i \epsilon_{ijk} X^j \, \delta_\varepsilon (X^0)\partial_\alpha X^0  \, \partial_\beta X^k \right) \, , \nonumber \\
\label{vertexangmom}
\end{eqnarray}
where we have again applied the two-dimensional Stokes theorem, $\alpha, \beta=1,2$ 
are world-sheet indices, and $\epsilon^{\alpha\beta}$ is the world-sheet Levi-Civita tensor, and
$\epsilon_{ijk}$ is the antisymmetric symbol in the three spatial dimensions of the brane world.
The logarithmic conformal properties~\cite{recoil} of the deformation arise from the $X^j$ parts.  
For the relatively large times $X^0 > 0$ after the impulse that we consider here, we may ignore 
the $\delta$-function terms, and in this case the effects of the angular momentum deformation 
in target space are equivalent to the open string propagating in an antisymmetric tensor ``magnetic-field'' 
type background with spatial components given by
\begin{equation}\label{magfield}
T^{-1} B_{ij} = - T^{-1} B_{ji} = \epsilon_{ijk} u^k~.
\end{equation}
This should be combined with (\ref{Bfoam}) in order to provide a complete description of the
averaged interactions of a photon with the D-foam, in a first-quantized version.

The quantity $u_{i}$ in (\ref{Bfoam}), 
which involves the momentum transfer, $\Delta p_{i}$, can be modelled by
a local operator using the following parametrization~\cite{sarkar}:
\begin{equation}
u_{i}=g_{s}\frac{\Delta p_{i}}{M_{s}}=\frac{g_s}{M_s} r_{i}p_{i}\,~,\, \quad {\rm no~sum~over}\, \quad i=1,2,3~,\label{defu2}\end{equation}
where the (dimensionless) variables $r_{i},i=1,2,3$, appearing above
are related to the fraction of momentum that is transferred in collision
with a D-particle in each spatial direction $i$. In the stochastic foam approximation~\cite{sarkar}
these parameters are taken as Gaussian normal random variables with a range $-\infty$
to $+\infty$ and defining moments 
\begin{equation}
<r_{i}>=0,\label{moment1}
\end{equation}
 \begin{equation}
<r_{i}r_{j}>=0, \quad \textrm{if}\quad i\neq j\label{moment2}
\end{equation}
 and 
 \begin{equation}
\sigma_{i}^{2}=<r_{i}^{2}>-<r_{i}>^{2}=<r_{i}^{2}>\neq0 \, .
\label{moment3}
\end{equation}
We assume that the foam is isotropic, which implies that $r_{i}=r$ for all $i=1,2,3$.
In this case the variances
\begin{equation}
\label{varianceisotr}
<(r_{i})^{2}>=\sigma^{2}~, \qquad i=1,2,3~,
\end{equation} 
are equal in all spatial directions.

However, for full rotational invariance in three space of the Born-Infeld action (\ref{bi}),
involving interactions between the velocity fields $u_i$ and the Maxwell tensor $F^{\mu\nu}$ of the vector field $\mathcal A_\mu$,
one more requirement is necessary. When considering
the application of the above averages to the recoil velocities, the latter must take the form:
\bea\label{gaussian2}
<u_i > = 0~, \qquad
  < u_i u_j > = \delta_{ij} \frac{\sigma^2\,g_s^2}{M_s^2} p^k p_k
\eea
The averages $< \dots >$ denote both statistical averages over populations of D-particle defects in the foam, 
\emph{cf.} Fig.~\ref{fig:dfoam}, as well as target-space quantum fluctuations. The latter can be induced by 
considering the summation over world-sheet genera~\cite{szabo}.
This is quite important for our purposes, and we return later to this issue and its implications.

For the moment, we remark that the stochasticity conditions (\ref{moment1})  imply
the restoration of Lorentz invariance in the statistical mean, with non-trivial fluctuations 
described in the isotropic and three-space rotationally-invariant case by (\ref{moment2}), 
(\ref{varianceisotr}) and (\ref{gaussian2}). Hence, although Lorentz invariance and spatial
isotropy is lost locally in individual scatterings of photons with D-particles,
due to the presence of the recoil velocity of the D-particle, $u_i$, the isotropy of the foam
washes such violations out on average, and isotropy and rotational invariance are restored.

\subsection{Finsler-Born-Infeld (FBI) Effective Actions}

We turn now to the singlet fermion sector. Unlike the charged fermions of~\cite{mavrolorentz}, 
here in principle there are recoil-induced contributions, since a singlet fermion can interact directly 
with the D-brane, causing recoil of the latter. Since the recoil is described by the antisymmetric tensor 
field excitations $B_{\mu\nu}$ in a $\sigma$-model framework, string theory gives specific rules for 
coupling this background to fermionic excitations propagating in the bulk. Any bulk world-sheet action
would involve the field strength of $B_{\mu\nu}$, namely $H_{\mu\nu\rho} = \partial_{[\mu}\, B_{\nu\rho]}$, 
where $[\dots ]$ denotes total antisymmetrization of the indices. Such field strength terms are dictated by 
an Abelian gauge symmetry $B_{\mu\nu} \to B_{\mu\nu} + \partial_{[\mu}\, \Lambda_{\nu]}$ that 
characterises the closed-string sector describing the gravitational multiplet of the string, to 
which the antisymmetric tensor belongs. A minimal string theory coupling between fermions and the
$H$-field strength would be of the form~\cite{string}: 
\begin{equation}\label{hterm}
\int d^4 x \epsilon_{\mu\nu\rho\sigma}\, H^{\nu\rho\sigma} \, \overline \psi \gamma^\mu \, \gamma^5 \, \psi~, 
\end{equation}
where $\epsilon_{\mu\nu\rho\sigma}$ is the covariant Levi-Civita tensor and $\gamma^5 \, \psi = \psi$
for right-handed fermions $\psi$. In principle, a non-zero 
$H$ field strength could arise in our model from inhomogeneous foam situations, in which
the variance (\ref{varianceisotr})  would depend on space-time. In the approximation of a 
homogeneous foam that is appropriate in the current epoch of our universe that is of interest
in the present article, we can neglect any such dependence on the 
variances (\ref{gaussian2}), and thus set to zero couplings of the form (\ref{hterm}). 

There will also be quantum fluctuations of the D-particles that are independent of 
their interactions with stringy matter (fermions or photons) and hence of the consequent recoil of the 
D-particle. Following the example of phonons in solids, which describe the quantum vibrational modes 
of the lattice ions, we represent the quantum fluctuations of the D-particles by vector Abelian fields, 
${\mathcal A}_\mu$, which are represented in string language by open strings with their ends 
attached to the D-particle. Being themselves D-branes, D3-particles can emit such open strings, 
which thus propagate in the bulk space between two D-particles in the foam. If the D-particles 
are confined to a single brane world, the field strength of the vector ${\mathcal A}_\mu$, 
which represents the intensity of the low-energy wave emitted by the quantum-fluctuating D-particle,  
is confined on the brane world, otherwise it propagates in the higher-dimensional bulk. 
In both cases the open string, whose low-energy vibrational modes contain ${\mathcal A}_\mu$, 
has Neumann boundary conditions over the entire available space. This picture is equivalent to the 
solid-state picture of phonons, which, though they owe their existence to vibrations of lattice ions,
are not confined to the position of the ion but travel through the entire lattice, interacting with the 
propagating electrons.

The field $\mathcal A_\mu$ plays the r\^ole of a massless Goldstone boson,
which is of vector nature as a consequence of the spontaneous breaking of
translational invariance by the presence of the D-particles in the D-foam vacuum~\cite{dfoam}. 
In Lorentz-violating field theories, with spontaneous breaking of Lorentz symmetry, 
the emergence of massless vector Goldstone bosons has been noted in the past, 
and attempts have been made to identify the photon with such a vector Goldstone boson~\cite{tomboulis}. 

We must stress that the vector fields ${\mathcal A}_\mu$  should  not be confused with ordinary 
photons $A_\mu$, as discussed in \cite{mavrolorentz}, which are gauge fields of the electromagnetic 
U(1) symmetry in the Standard Model. However, in a similar way to phonons interacting with electrons in a solid, the vector 
fields ${\mathcal A}_\mu$ interact with singlet fermions, and are represented in string language by 
open strings stretching between the D-particle and the D3-brane world. As in the photon-D-particle
interaction case discussed above, the recoil of the D-particle during its interaction with the singlet fermion
is represented by an antisymmetric tensor field, whereas the quantum vibrations of the D-particle lattice 
are represented by the vector field ${\mathcal A}_{\mu}$ that lives in the entire bulk space. In addition, 
there is an explicit coupling $\tilde g_V$  of the field ${\mathcal A}_\mu$ to the fermion current, 
similar to the photon-fermion current of charged fermions in QED. However, the vector field
${\mathcal A}_\mu$ does not have an interpretation as a conventional gauge field.

We have seen that in homogeneous foam situations the only coupling of the singlet fermion
would be through the vector field ${\mathcal A}_\mu$:
\begin{equation}\label{fermion}
S_{\psi} = \int d^4 x {\overline \psi} \gamma^\mu i\,D_\mu \psi~, \quad D_\mu = \partial_\mu + i {\tilde g}_V \, {\mathcal A}_\mu~.
\end{equation}
where we repeat that ${\mathcal A}_\mu$ is not the photon but the vector field representing the quantum fluctuations 
of the D-particles, which was absent in the charged fermion case. We recall that in the charged-fermion 
case of \cite{mavrolorentz} there was also no direct coupling of the charged fermion to the recoil terms, 
but for a different reason. There, such couplings were identically zero because of the properties of the 
D-foam, according to which the latter is transparent to charged-fermion fields, because of charge 
conservation. This argument is strictly speaking valid only for Type IIA string theory D-foam, 
in which the D-particles are point-like. For type IIB string theory D-foam models, on the other hand, 
D-particles are compactified D3-branes, and as such there are non-trivial, but much more
suppressed, fermion-foam couplings~\cite{li}. Hence, also in that case the lowest-order 
(weak-field) effective action term in the fermion sector is given by the QED-like fermion-vector
coupling (\ref{fermion}), where however the photon field of QED is replaced by the vector field ${\mathcal A}_\mu$. 

The effective target-space action on the D3-brane world, where the quantum fluctuating D-particle 
meets the open-string singlet fermion state, is described by the following Born-Infeld Lagrangian~\cite{seibergwitten,sussk1} 
augmented with the interaction of the fermion current with the vector field ${\mathcal A}_\mu$:
\begin{equation} \label{bi}
S_{BI} = \frac{T^2}{g_s} \int d^4 x \sqrt{{\rm det}\left(g + T^{-1} (B + F)\right)} + \int d^4 x \, {\tilde g_V} \, {\mathcal A}_\mu \overline \psi \gamma^\mu \, \psi  + \int d^4 x \, \overline \psi i \, \slash{\partial} \psi~,
\end{equation}
where $F_{\mu\nu}=\partial_\mu {\mathcal A}_\nu - \partial_\nu {\mathcal A}_\mu$ is the field strength tensor of the 
vector field ${\mathcal A}_\mu$, and the fermion field $\psi$ represents a massless Majorana spinor, and the coupling 
${\tilde g}_V $ is discussed below. We remark at this stage that, if photons were to be considered,
then an additional term involving the Maxwell field-strength  for the ordinary photons 
would be present in the argument of the square root of the first term of the action (\ref{bi}).

The Lagrangian (\ref{bi}) depends on both space-time coordinates and the momentum transfer, 
and hence cannot be expressed as an ordinary local term in an effective action framework.
Nevertheless, as was argued in \cite{mavrolorentz}, in the 
approximation (\ref{defu2}), in which the
D-foam background is treated stochastically, the construction of a low-energy local 
effective action for the fields ${\mathcal A}_\mu$ and $\Psi$ becomes possible, and that action 
resembles the Lorentz-violating theory appearing in \cite{alexandre}.

Let us now return to the effective action (\ref{bi}). Defining the generalized field 
$\mathcal{F}_{\mu\nu} \equiv F_{\mu\nu} + B_{\mu\nu}~$, we make use of the fact~\cite{tseytlin}
that in four Minkowski space-time dimensions the determinant
${\rm det}_4 \left(\eta + T^{-1}\,F\right)$ has special properties that allow the following 
representation of the Born-Infeld action (we work from now on in units where the string tension 
is $T=1$)~\footnote{We remind the reader that the indices are raised and lowered with respect to
the background metric $g$, which we assumed to be Minkowski $\eta$. In a general situation, 
where the metric $g$ is not trivial, the pure foam contributions proportional to various powers 
of $u_i^2$ contribute to (dark) vacuum energy~\cite{westmuckett,emnnewuncert}, and can be 
constrained by cosmological considerations, for instance. We ignore such terms in our discussion here.}:
\bea\label{4dbi}
S_{BI} & = &  \frac{1}{g_s} \, \int d^4 x \left(I_2 + I_4 \left(1 + \mathcal{O}(\mathcal{F}^2) \right)\right) +{\rm const}
\nonumber \\
\quad I_2  &=& \frac{1}{4}\mathcal{F}_{\mu\nu}\mathcal{F}^{\mu\nu}~, \quad I_4 = -\frac{1}{8} \left[\mathcal{F}_{\mu\nu}\,\mathcal{F}^{\nu\rho} \, \mathcal{F}_{\rho\lambda}\,\mathcal{F}^{\lambda\mu} -
\frac{1}{4}\left(\mathcal{F}_{\mu\nu}\mathcal{F}^{\mu\nu}\right)^2\right]~.
\eea
In the weak vector field approximation of interest to us here we can ignore terms of order higher than 
quadratic in the vector field and the (small) recoil velocity $u_i$ field. This is a consistent approximation 
for relatively heavy D-particles, whose recoil is suppressed by their mass.
We also take a quantum average over stochastic fluctuations of the $B$-field, using (\ref{gaussian2}), 
and keep terms quadratic in the vector (${\mathcal A}$) or (averaged) recoil ($<u>$) fields, including mixed terms of order ${\mathcal A}^2 <u^2>$.

\subsection{Target-Space Quantization of the FBI Action}

The summation over world-sheet genera leads to target-space quantization of the 
Finsler background $B_{0i}$ (\ref{Bfoam}), as discussed in detail in~\cite{szabo}.
The properties of the D-particle recoil vertex operator as a (logarithmic) conformal field theory in a $\sigma$-model approach~\cite{recoil}
lead, upon summation over world-sheet topologies, to quantum uncertainty relations between the 
the recoil velocities and the collective coordinates describing the initial position of the D-particle,
which correspond to those of canonically-quantized 
collective momentum and position operators for the D-particle in target space. It was therefore argued 
in~\cite{szabo} that, via (\ref{defu2}), the summation over world-sheet genera elevates the recoil velocity fields 
$u_i$ into quantum-fluctuating momentum operators in target space~\footnote{The effects of the summation 
over genera on the fluctuations of the 
background fields $u_i$ have been expressed in closed form only in the bosonic string case. Although
closed expressions have not been derived in the world-sheet supersymmetric case~\cite{szabosusy},
we expect our arguments on the correspondence principle (\ref{quant})
to characterize all types of string models.}:
\bea\label{quant}
B_{i0} = u_i  \Rightarrow  \widehat{B}_{i0} = \widehat{u}_i = -i g_s \, \frac{r_i}{M_s} \hbar \frac{\partial}{\partial X^i} &\equiv & -i g_s \, \frac{r_i}{M_s} \hbar \nabla_i~,
\;  ({\rm no~sum~over~} i = 1,2,3) ~. \nonumber \\
 B_{ij} = \epsilon_{ijk}\,u^k & \Rightarrow &  \widehat{B}_{ij} = \epsilon_{ijk}\,\widehat{u}^k~.
\eea
Consequently, the uncertainty relations $[ X^i, t ] \sim u^i$ of the classical recoil 
background~\cite{review} are also elevated to quantum operator relations~\footnote{The correspondence (\ref{quant}) 
also leads to master equations that can be used to study the induced 
decoherence of quantum matter propagating in a quantum-fluctuating D-foam background~\cite{sarkar}.}.

There is an important aspect of the correspondence (\ref{quant})
that we use in the following. This 
correspondence was derived from the $\sigma$-model approach to recoil~\cite{recoil}, 
in which one is forced to use time fields $X^0$ with Euclidean signature
in order to guarantee the convergence of the world-sheet path integrals. This leads us to extend the correspondence (\ref{quant})
to include~\cite{mavrolorentz}:
\bea\label{eucl}
&& B^{i0} \Rightarrow \widehat{B}^{i0} =   g^{00}_E g^{ik}_E\widehat{B}_{k0} = + \widehat{B}_{i0}~,
\eea
where the subscript $E$ indicates the Euclidean signature. We revert to 
Minkowski signature by analytic continuation only at the end of our computations, 
after replacing the background $B$ fields by appropriate operators.

In this correspondence (\ref{quant}, \ref{eucl}), the statistical fluctuations in the foam
are implemented by averaging statistically the momentum transfer variable $r$ ($<...>$) over the
population of quantum-fluctuating D-particles, using the relations (\ref{gaussian2}) for the case of isotropic foam,
leading to the correspondence:
\bea\label{gaussian}
<u_i > = 0~, \quad
  < u_i u_j > = \delta_{ij}\, \frac{\sigma^2\,g_s^2}{M_s^2} p^k p_k \quad \Rightarrow \quad - \hbar^2 \frac{\sigma^2\, g_s^2}{M_s^2} \delta_{ij} \Delta~, \; \Delta \equiv {\vec \nabla} \cdot {\vec \nabla}~.
\eea
Via this prescription for first quantization, the effective action (\ref{4dbi}) is mapped
onto a particularly simple local effective action, as we now show.

\subsection{A Minimal Lorentz-Violating QED-Like Action}

After implementing the correspondences (\ref{quant}, \ref{eucl}), 
we obtain the following vector-field-dependent terms in an action of Finsler-Born-Infeld (FBI) type:
\begin{eqnarray} \label{bioper}
&&S_{BI} \ni \frac{1}{g_s} \,\int d^4 x  : \left[\frac{1}{4} F_{\mu\nu}\left(1 +
\frac{b}{16}\,\widehat{B}_{\alpha\beta}\widehat{B}^{\alpha\beta}\right)F^{\mu\nu}
+ \frac{a}{64} F_{\mu\nu}\,\widehat{B}^{\mu\nu}\widehat{B}_{\alpha\beta}\,F^{\alpha\beta} \right.
\nonumber \\
&&\left. -\frac{1}{8}\,F_{\mu\nu} \widehat{B}^{\nu\rho} \widehat{B}_{\rho\beta} F^{\beta\mu}\right] : + \dots \, ,
\end{eqnarray}
where the $\dots$ in (\ref{bioper}) represent terms of higher order in the fields $B$ and $F$, and the symbol : \dots : 
denotes an appropriate quantum operator ordering, in which the ordering constants $a, b$ obey~\cite{mavrolorentz}
\be\label{ordering}
a + b =2~.
\ee
Terms involving odd powers of the operators $\widehat{B}_{\mu\nu}$,
vanish in our stochastic Gaussian background (\ref{gaussian}), and so may be ignored. 
Terms in which the operators $\widehat{B}_{0i}$ or $\widehat{B}_{ij}$
lie on the far left-hand-side of the integrand in (\ref{bioper}) have also been dropped, as they correspond to 
total spatial derivative terms that do not contribute, under the conventional assumption that the fields decay suitably
at spatial infinity on the brane world.

Using as an intermediate step a target-space time with Euclidean signature when making the correspondence (\ref{gaussian}), 
as discussed above (see (\ref{eucl})), some straightforward algebra using (\ref{ordering}) reduces (\ref{bioper}) 
to~\cite{mavrolorentz}
\begin{eqnarray}\label{bioperfinal}
S_{BI} \ni \frac{1}{g_s} \,\int d^4 x   \left[\frac{1}{4} F_{\mu\nu}\left(1
 + \frac{1}{4}\,(1 - \frac{b}{2})\, \frac{g_s^2 \sigma^2}{\,M_s^2} \Delta \right) \,F^{\mu\nu} \right] \, + \, \dots ~,
\end{eqnarray}
where the $\dots$ represent terms of higher order in derivatives and the Maxwell tensor $F_{\mu\nu}$, 
and $\Delta$ is the 3-space Laplacian, $\Delta \equiv \nabla_i \nabla^i = \vec{\nabla} \cdot \vec{\nabla}$.

Whereas the recovery of the FBI action is trivial in the classical limit, quantum ordering ambiguities are an issue. 
Usually, the quantum ordering of operators is specified by requiring that the effective Lagrangian be Hermitian, 
which is not an issue here because of the stochasticity of the foamy background (\ref{gaussian}). 
In principle, is selected by  the correct quantum ordering is selected by the full underlying theory of quantum gravity, 
but this is still elusive, so we proceed phenomenologically.
As discussed in detail in \cite{mavrolorentz}, the choice $b=2$ would eliminate Lorentz-violating
terms, but a solution that leads to dynamical mass generation is preferred, in order to
avoid infrared (IR) divergences and the associated instabilities. Assuming that the 
full quantum gravity theory should act as an IR regulator, one must select an ordering with $b \ne 2$.

Any ordering with $b < 2$ would lead to terms that affecting the 
pole structure of the vector propagator in three-space, as the bare propagator of the vector boson ${\mathcal A}_\mu$
stemming from (\ref{bioperfinal}) has the following form (apart from gauge-fixing terms that we 
do not write explicitly here)~\cite{alexandre}:
\be\label{D2}
D_{\mu\nu}^{bare}(\omega,\vec p)=-\frac{i}{1  - p^2/M^2}\left( \frac{\eta_{\mu\nu}}{-\omega^2+p^2} -\frac{p_\mu p_\nu}{(-\omega^2+p^2)^2}
 \right) , \,\,   M^2 \equiv \frac{4\,M_s^2}{(1-b/2)\,g_s^2 \, \sigma^2} > 0~.
\ee
In contrast to the case of a conventional gauge field, we recall that ${\mathcal A}_\mu$ is a vector field without any gauge symmetry,
so there is no gauge ambiguity of the type encountered for photons~\cite{mavrolorentz}. Rather,
the propagator (\ref{D2}) corresponds to the gauge $\xi=0$ in the photon propagator of~\cite{alexandre}.

The effective action in Fourier space would therefore no longer be unitary at 
momenta above the  scale $M$. This is exactly what could happen~\cite{emnnewuncert,review} 
if the classical recoil velocity (\ref{recvel}) were to exceed the speed of light in vacuo, which is
unphysical. The low-energy effective action is completely consistent in a classical background,
and the mass scale $M$ defines the range of validity of the 
low-energy local effective action.

In extending our prescription beyond such an classical effective field theory,we look
for a choice of quantum ordering that allows the action to be extended beyond the classical limit,
so as to describe some aspects of space-time foam for
larger momenta. This would provide a partial ultraviolet (UV) completion of the low-energy theory
as far as dynamical mass generation is concerned~\footnote{We note that other aspects of the foam, 
such as vacuum refraction-induced photon delays, cannot be described within the framework of local 
effective field theories, see~\cite{emnnewuncert,review}.}. Dynamical mass generation is, however, 
an IR phenomenon and the detailed UV structure of the theory should not affect it. 
This is, for instance, what happens with the Landau pole of QED, 
whose presence does not affect dynamical mass generation~\footnote{There is support for this 
from lattice calculations but, as far as we are aware, no rigorous proof exists as yet~\cite{gms,lpole}.}. 
However, our case is different as the effective theory breaks down above a momentum scale where
the effective Lagrangian  is no longer unitary, because of a change in the sign of the photon propagator. 
A study of dynamical generation in such a case would need to cut the momentum integrals in the Schwinger-Dyson
equations off above the scale $M$. This would be different from the 
consistent Schwinger-Dyson treatment in~\cite{alexandre}, which required the cancellation of 
potential UV divergences, implying implicitly a suitable extension of the model beyond any UV cut-off.

Fortunately, it is possible to avoid such a UV cut-off with a suitable choice of quantum ordering.
A minimal class of such orderings, which yield a unitary photon propagator 
(\ref{D2}) in three space dimensions, is that in which the ordering parameter $b$ is in the range:
\be
b > 2 \, .
\label{physord}
\ee
After a formal analytic continuation back to Minkowski space-time, the action
(\ref{bioperfinal}) takes the same form (\ref{bare}) as in~\cite{alexandre}, with the mass scale
\begin{equation}\label{MMs}
M = \frac{M_s}{\,g_s \,\sqrt{{\tilde \sigma}^2}}~, \qquad \tilde{\sigma}^2 \equiv \sigma^2 \frac{1}{4}|1 - \frac{b}{2}|~.
\end{equation}
Any residual ordering ambiguity can be absorbed along with the fluctuations of the foam
into a small phenomenological parameter in our first-quantized approach. 
The remaining ambiguities should be removed
when a full, second-quantized quantum gravity model for D-foam fluctuations is developed 
~\footnote{It is interesting to note that for the unique
value $b=10$, $\tilde{\sigma}^2 = \sigma^2 $, so that the scale $M$ of the Lorentz-violating
terms in the action (\ref{bioperfinal}) is identical to that at which the foam-averaged 
recoil velocity equals the speed of light in vacuo~\cite{emnnewuncert,review}. In this case
a unique quantum gravity scale would enter the model in different guises.}. 
After such a suitable choice of quantum ordering the effective action exhibits
maximal suppression of the Lorentz-violating effects and cures IR instabilities through dynamical mass generation.

As already mentioned above and emphasised in~\cite{review,emnnewuncert}, 
the phenomenon of induced vacuum refraction for photons cannot be captured by this local effective action. 
The time delays of photons that are induced by their topologically non-trivial interactions with the D-foam are expected to
scale linearly with the photon energy, and thus are suppressed by just a single power of the string mass scale, 
are purely stringy effects, associated with time-space uncertainties~\cite{sussk1,emnnewuncert,li,review} 
and generated by intermediate (non-local) string states stretched between the the D3 branes and the D-particles. 
This is to be contrasted with the modifications due to the foam in the local string effective action (\ref{bioperfinal})
or (\ref{bi}), which are quadratically suppressed by the string scale (\ref{MMs}), as a result of the 
stochasticity assumption (\ref{gaussian}).

We now discuss dynamical mass generation for the singlet fermions as a consequence 
of the dynamics described by the local effective action (\ref{bi}).

\section{Dynamical Mass Generation for Singlet Fermions \label{sec:3}}

In ref. \cite{alexandre}, dynamical mass generation for fermions has been studied in the 
context of a (3+1)-dimensional QED-like field theory with higher-order spatial derivatives in the photon sector
that violated four-dimensional Lorentz symmetry but preserved spatial rotations, and a standard form for the fermion  sector.
This Lorentz-violating model is not of Lifshitz type~\cite{horava}, in the sense that there is isotropic scaling between 
time and space coordinates, but there is a mass scale that suppresses the Lorentz-violating spatial-derivative terms,
as we now describe.

\subsection{Review of a Minimal Lorentz-Violating QED-Like Model \label{sec:2}}

The Lorentz-violating Lagrangian considered in \cite{alexandre} is:
\be\label{bare}
{\cal L}=-\frac{1}{4}F^{\mu\nu}\left(1-\frac{\Delta}{M^2}\right)F_{\mu\nu}
-\frac{\xi}{2}\partial_\mu A^\mu\left(1-\frac{\Delta}{M^2}\right)\partial_\nu A^\nu
+i\ol\psi\br D\psi \, ,
\ee
where $\xi$ is a covariant gauge-fixing parameter, $D_\mu \equiv \partial_\mu+ieA_\mu$, 
and $\Delta \equiv \partial_i\partial^i=\vec\partial\cdot\vec\partial$, and in our convention the
metric  is (-1, 1, 1, 1). As explained in the previous Section, our non-gauged model action
describing the low-energy interactions of singlet fermions $\psi$ with quantum-fluctuating D-foam
consists of the terms (\ref{bioperfinal}) and (\ref{fermion}), 
and corresponds to (\ref{bare}) after fixing the gauge $\xi=0$.

There were no higher-order spatial derivatives for the fermions in \cite{alexandre}, 
so as to avoid non-renormalizable couplings in the theory of the form
\be\label{fermiho}
\frac{1}{M^{n-1}}\ol\psi (i\vec D\cdot\vec\gamma)^n\psi~~~~~~n\ge 2 \, ,
\ee
and such terms are absent in the framework we introduced in the previous Section.
Standard (3+1)-dimensional QED in a covariant gauge is recovered in the limit $M\to\infty$, and
this scale characterizes the energies at which Lorentz-violating effects become important. 
It may be the Planck scale, or not, depending on the microscopic origin of Lorentz violation.

The Lorentz-violating terms in (\ref{fermiho}) play a dual r\^ole, as discussed in \cite{alexandre}:

\begin{itemize}
\item First, they introduce a mass scale, $M$, that is needed for generating a fermion mass. In
the model of~\cite{alexandre}
\be\label{dynmass}
m_{dyn} = M {\rm exp}\left(-\frac{2\pi}{\left(4 + (\xi - 1)\right)\alpha}\right)~,
\ee
where $\alpha = e^2/4\pi$ is the fine structure constant.
There is an analogy with the magnetic catalysis phenomenon of standard QED~\cite{B,gms}, according to which a 
sufficiently strong magnetic field catalyzes the dynamical generation of a fermion mass for arbitrarily weak QED couplings. 
This is an example of Lorentz violation, with the Lorentz symmetry breaking being provided 
by the direction of the background magnetic field. However, in our model there are two important differences 
from that of \cite{alexandre}. One is that the magnetic field breaks three-dimensional rotational symmetry and
induces an effective dimensional reduction to two dimensions, as a result of the 
(1+1)-dimensional form of the fermion propagator in the lowest Landau level, 
which is dominant in the strong magnetic field case.

\item Secondly, the higher-derivative Lorentz-violating terms
provide an effective regularization of the theory, leading to finite gap equations~\cite{alexandre}.
We emphasize, in order to avoid confusion, that the scale
$M$ appearing in this approach does \emph{not} regulate the theory (\ref{bare}), as it regularizes loops
with an internal photon line only. Rather, $M$ is a parameter of the model upon which depend physical quantities
such as the dynamically-generated mass. As we have seen in Section \ref{sec:dfoam}, 
in our stringy quantum-gravity model that is described by the Lagrangian (\ref{bare}) in the 
low-energy field-theoretical limit, this scale is expressible in terms of fundamental parameters 
of the underlying string theory, see (\ref{MMs}). 
\end{itemize}

The following bare photon propagator was derived from the Lagrangian (\ref{bare}) in~\cite{alexandre}:
\be\label{D}
D_{\mu\nu}^{bare}(\omega,\vec p)=-\frac{i}{1+p^2/M^2}\left( \frac{\eta_{\mu\nu}}{-\omega^2+p^2} +
(\xi -1)\frac{p_\mu p_\nu}{(-\omega^2+p^2)^2}\right) \, ,
\ee
where $p^0= \omega$ and $p^2=\vec p\cdot\vec p$. Note that, since the 
pole structure is \emph{not} affected  by the Lorentz-violating terms, 
the photon remains \emph{massless} in this minimally Lorentz-violating model~\cite{alexandre}.
The dynamical mass (\ref{dynmass}) was derived from the following
Schwinger-Dyson equation for the fermion propagator (see, e.g.,~\cite{miransky}):
\be\label{SD}
S^{-1}-S_{bare}^{-1}=\int D_{\mu\nu}(e\gamma^\mu) S\,\Gamma^\nu \, ,
\ee
where $\Gamma^\nu$, and $S$ and $D_{\mu\nu}$ are the dressed vertex, the 
dressed fermion propagator and the dressed photon propagator, respectively. The equation (\ref{SD}) gives an exact
self-consistent relation between dressed $n$-point functions, and thus is \emph{non-perturbative}.
As a consequence, the would-be divergences are not absorbed by
redefinition of the bare parameters in the theory, but the equation is regularized by the scale $M$, 
which thereby acquires physical significance.

The Schwinger-Dyson equation (\ref{SD}) was solved in~\cite{alexandre} using the ladder approximation. 
In this approximation one ignores corrections to the vertex function, which would have led to a
system of coupled Schwinger-Dyson equations that would complicate matters significantly. 
This approximation is known not to be gauge invariant~\cite{miransky},
a problem that is generic in off-shell field-theoretic quantities that appear at intermediate stages in
calculations of physical on-shell quantities. There are some gauges, termed non-local gauges, 
in which this bare approximation to the vertex is argued to be an exact Ansatz~\cite{gms}. 
In our discussion below we restrict our analysis to one loop, and pick out the 
gauge-independent part of the dynamically-generated mass in a QED-like theory.
In this spirit, loop corrections to the photon propagator fermion wave-function renormalization
were neglected in~\cite{alexandre}, and only the corrections to the electron self-energy
were kept. 

With our approximations, the dressed fermion propagator can be expressed as
\be\label{G}
S(\omega,\vec p)=i \,\frac{ p_\mu \gamma^\mu - m_{dyn}}{p_\mu p^\mu + m_{dyn}^2 } \, ,
\ee
where $m_{dyn}$ is the fermion dynamical mass,
and the Schwinger-Dyson equation (\ref{SD}), which involves a convergent integral
thanks to the $M$-dependent Lorentz-violating terms, becomes
\be
m_{dyn}=\frac{\alpha}{\pi^2}\int \frac{d\omega ~p^2dp}{1+p^2/M^2}\frac{m_{dyn}(3+\xi)}{(-\omega^2+p^2)(-\omega^2+p^2+m^2_{dyn})} \, ,
\ee
where the fine structure constant $\alpha=e^2/4\pi$.
This equation has the obvious solution $m_{dyn}=0$, but we are interested in a potential second solution, which
satisfies the following gap equation that is obtained after integration over the frequency $\omega$:
\be\label{gap}
\frac{\pi}{(3 +\xi)\alpha}=\int_0^\infty\frac{x\,dx}{1+\mu^2x^2}\left( 1-\frac{x}{\sqrt{1+x^2}}\right) \, ,
\ee
where $\mu \equiv m_{dyn}/M$ is the dimensionless dynamical mass, which we assume to be small.
Both of the terms in parentheses in (\ref{gap}) are divergent, but these divergences cancel when combined. 
As explained in \cite{alexandre}, after some approximations valid in the limit $\mu \ll 1$, 
the non-trivial dynamical fermion mass (\ref{dynmass}) arises as a consistent solution of the equation (\ref{gap}). 
The physical solution of the Schwinger-Dyson equation (\ref{SD}) is that with the dynamical mass 
$m_{dyn} \ne 0$, as it avoids IR instabilities.
It is worth noting that the expression (\ref{dynmass}) for $m_{dyn}$ is not analytic in $\alpha$,
and so can only be found in a non-perturbative approach,
such as the Schwinger-Dyson equation used in \cite{alexandre}.

As mentioned above and discussed in~\cite{mavrolorentz}, in a QED-like theory
where the vector field is a gauge field such as the photon and the fermion is charged like
the electron, this non-perturbative solution depends on the {gauge-fixing parameter} $\xi$.
However, in our case, where the vector field ${\mathcal A}_\mu $ 
does not have a gauge symmetry, there is no such gauge ambiguity,
and the model corresponds to the case with a fixed gauge parameter  $\xi =0$.

Thus, in our case, the physical mass of the singlet fermion generated dynamicallly by its interactions with the 
foam is: 
\bea\label{dynmass2}
m_{dyn}& =& M {\rm exp}\left(-\frac{2\pi}{3\,\alpha_V}\right)~, \nonumber \\ \alpha_V &=& \frac{{\tilde g}_V^2}{4\pi}~, \quad M = \frac{M_s}{\,g_s \,\sqrt{{\tilde \sigma}^2}}~, \quad \tilde{\sigma}^2 \equiv \sigma^2 \frac{1}{4}|1 - \frac{b}{2}|~, \, b \ne 2~,
\eea
where we have replaced the coupling $e$ of QED by the vector coupling ${\tilde g}_V$ (in \ref{fermion}), 
and the scale $M$ is given by (\ref{MMs}).
The magnitude of the coupling ${\tilde g}_V$ depends on the string model, and
we treat it here as a phenomenological parameter whose estimation we leave to future work.

\subsection{Application to Singlet Fermions}

If the singlet fermions are right-handed neutrinos, we expect them to be Majorana. Dynamical mass generation
for such neutrinos (whose left-handed counterparts belong to SU(2) doublets of the Standard Model)
in the above minimal Lorentz-violating framework has been discussed in detail in \cite{leite}, 
whose results we borrow for our problem here. As discussed in \cite{leite}, the relevant low-energy 
Lorentz-violating Lagrangian is:
\bea\label{Lag3}
\mathcal{L}&=& -\frac{1}{4} F_{\mu\nu}(1-\frac{\Delta}{M^2})F^{\mu\nu}+ 
\overline{N}(i \slashed{\partial}- \tilde g_V \slashed{A})\, \frac{1}{2} \Big(1 + \gamma_5 \Big)\, N \\
&&~~~~~~~~~~~~~~~~~+\overline{\nu}(i \slashed{\partial}- e_2 \slashed{A})\, \frac{1}{2} \Big(1 - \gamma_5 \Big)\, \nu~,\nonumber
\eea
where we have concentrated on the interactions of the neutrinos with the foam, 
ignoring the SU(2) weak interactions of $\nu_L$ for brevity and concreteness. 
The notations in (\ref{Lag3}) are the following: the Majorana right-handed neutrinos are given by 
$\nu_R \equiv \frac{1}{2} \Big(1 + \gamma_5 \Big)\, N$ and the left-handed neutrinos of the Standard Model sector, 
belonging to SU(2) doublets with the ordinary charged leptons, by $\nu_L = \frac{1}{2} \Big(1 - \gamma_5 \Big)\, \nu$, 
where $N$ and $\nu$ are four component non-chiral Majorana spinors, with zero bare mass, 
and for simplicity we restrict ourselves to one generation only for now, commenting later on realistic models with three generations. 
The coupling $e_2$ is either zero (in point-like D-foam models~\cite{dfoam,emnnewuncert} ) or strongly suppressed 
compared to $\tilde g_V$ (for type IIB string D-foam models~\cite{li}), by factors that depend on the geometrical characteristics
of the foam.

As an example, we consider the type IIB model of \cite{li}, in which
our world is viewed as a D7 brane with four dimensions compactified, 
and the ``D-particles'' of the foam are represented by compactified D3 branes wrapped around three cycles. 
Assuming that there is one D-particle per three-volume $V_{A3}$,  and denoting
the radius of the fourth space dimension of the D7 brane transverse to the D3 brane by $R^\prime$, 
this suppression factor is given by
\begin{equation}
\eta \equiv \frac{e_2^2}{\tilde g_V^2} \;  = \; \frac{(1.55\, \ell_s)^4 \, n_D^{(3)}}{R^\prime} \, , 
\end{equation}
where $n_D^{(3)} = V_{A3}^{-1}$ is the D-particle-foam density in three-space. 
In the model of~\cite{li}, the values $V_{A3} \sim (10\ell_s)^3$ and 
$R^\prime \sim 338 \ell_s$ were shown to be consistent with phenomenology,  
but different values are possible in other models. An order of magnitude estimate of $\tilde g_V$ 
can be given in that model by noticing~\cite{li} that the gauge coupling $g_{37}$ describing interactions of 
particles on the D7 brane (including singlet fermions) with D-particles (compactified D3 branes) is
\begin{equation}
\label{gauge}
g_{37}^{-2} \; = \; V \, g_7^{-2}~,
\end{equation}
where $V \equiv V_{A3}\, R^\prime$ is the volume of the extra four spatial dimensions of D7 branes 
transverse to the D3 branes, and the coupling $g_7^2 \propto g_s$, where $g_s$ is the string coupling. 
The coupling $g_{37}$ (\ref{gauge}) is identified with the $\tilde g_V$ coupling in case of singlet fermion excitations,
so we have
\bea\label{form}
\tilde g_V \propto g_s^{1/2} \sqrt{n_D^{(3)} \, {R^\prime}^{-1}} \, {\mathcal F}(s,t, \alpha^\prime)~,
\eea
where $s$ and $t$ are appropriate $s$- or $t$-channel Mandelstam variables in momentum space
and ${\mathcal F}(s,t, \alpha^\prime)$ is a momentum-dependent form factor associated with string 
amplitudes describing the scattering of such singlet fermionic excitations off D-particles in the model, 
including string-loop corrections that are suppressed by powers of $g_s$
which ``renormalises'' the coupling $\tilde g_V$.  This form factor is difficult to compute exactly,  given 
that the target-space action of D-branes is not fully known. For slowly-moving excitations 
with momenta that are small compared to the string mass scale, $M_s=1/\sqrt{\alpha^\prime}=1/\ell_s$,
for which a field-theoretical approximation is adequiate, the form factor is well approximated by unity.
However, for higher momenta, and over a wide range of energies, stringy effects of order $\alpha^\prime$
become comparable to field-theoretical contributions, as can be seen by rough estimates of 
form factors in concrete models by applying T-duality arguments, which map the complicated 
non-perturbative D-brane/string amplitudes into perturbative string-string scattering amplitudes~\cite{shiu}. 
It is apparent from this discussion that precise estimates of $\tilde g_V$ depend on the
microphysical model, in addition to being proportional to the density of the D-particles in the foam.  

Following the detailed discussion  in \cite{leite},  we may 
regard the initially massless $\nu$ and $N$ as a Majorana doublet $\nu^M = (\nu, N)$
that satisfies the Majorana condition $\nu^M = (\nu^M)^c$, with $c$ denoting the usual charge conjugation operation,
and couples to the vector fields. It was shown in~\cite{leite} by solving the corresponding system of coupled 
Schwinger-Dyson equations for the fermions $\nu$ and $N$ that, in the single gauge field toy models of relevance here, 
the only solution for dynamical fermion mass generation is one in which the mass eigenvalues $m_1,m_2$
are
\bea\label{dgm}
m_1&=&\lambda_-=0\\
m_2&=&\lambda_+=M\exp\left(\frac{-8\pi^2}{3\, {\tilde g}_V^2}\right)\nonumber~,
\eea
upon fixing the gauge parameter $\zeta = \xi -1 $ of \cite{leite} to $\zeta = -1 $. The mass
$m_1$ can be identified with the left-handed Majorana mass $M_L=0$,
and $m_2$ (which is identical to $m_{dyn}$ in (\ref{dynmass2}) is identified with the heavy right-handed 
Majorana mass $M_R$.

There is no non-trivial Dirac mass $\mu$ in the dynamical solution, as explained in \cite{leite},
but a Dirac mass term can be generated through the usual Yukawa coupling with the Higgs field. 
The (weak) foam coupling $\tilde g_V \ll 1 $ and the Yukawa coupling to the Higgs field in this scenario
can be chosen in such a way that that the right-handed neutrino Majorana mass (\ref{dynmass2})
is much heavier than the Higgs-generated Dirac mass, leading naturally to a light active neutrino
in the Standard Model sector, as in a conventional seesaw model. 
In the realistic case of more than one flavour of right-handed neutrino, e.g.,
three flavours to match the number of active neutrinos as in the $\nu MSM$ model~\cite{nuMSM}, 
one can arrange the active neutrino mass hierarchy by appropriate choice of the Yukawa couplings, 
even if there is a degeneracy of right-handed Majorana masses (\ref{dynmass2}), as
in models where there is a universal geometric coupling $\tilde g_V$. In other models, 
$\tilde g_V$ may be species-dependent, which could lead to a mass hierarchy between 
the right-handed neutrino Majorana masses.

Before closing this section we remark that it is possible to improve the ladder approximation 
so as to replace the bare coupling $\tilde g_V$ in (\ref{dynmass2}), (\ref{dgm})  
by a running one. This would modify the mass (\ref{dynmass2}), but such an
improved analyses would not affect significantly the order of magnitude of the non-perturbatively 
generated mass (\ref{dynmass2})~\footnote{The only case where such an analysis enhances
significantly the dynamical mass compared to the ladder approximation is magnetic 
catalysis~\cite{gms} in the presence of an external magnetic field. However, the reason 
why the dynamical mass is less suppressed in the improved approximation is the 
effective dimensional reduction to two dimensions induced by the magnetic field. 
There is no such a reduction in our Lorentz-violating case, since the singlet fermions are not charged.}. 
However, a geometric enhancement of the dynamical mass in multi-brane-world 
scenarios \`a la Randall-Sundrum~\cite{RS} was discussed in \cite{mavrolorentz}, and is reviewed in the next Section.

\section{Geometrical Enhancement of Dynamical
Fermion Masses in Multi-Brane-World Scenarios \label{sec:4}}

We now present a geometric mechanism that enhances the dynamical 
fermion mass (\ref{dynmass2}) by a suitable embedding of the model in a higher-dimensional set-up
involving brane worlds in a Randall and Sundrum (RS) warped bulk geometry~\cite{RS}. It was suggested in \cite{RS} that a
large hierarchy between the Planck mass and the TeV scale could arise in 
brane-world scenarios in which our world is a negative-tension brane located at a distance 
$r_c \pi$ from a hidden-sector brane embedded in a five-dimensional bulk. 

As in RS, we assume that the five-dimensional metric is a solution of 
Einstein's equations in an anti-de-Sitter bulk space with a warp factor of the form
\be
ds^2 = e^{-\sigma(z)} \eta_{\mu\nu} dx^\mu dx^\nu + dz^2 \, ,
\label{warp}
\ee
where $z$ is the bulk (fifth) dimension, and the $x^\mu$ are coordinates in our four-dimensional space-time.
Because of the warp factor $e^{-\sigma(z)}$ in the metric (\ref{warp}), a field on our brane world of mass $m_0$ 
with a canonically-normalized kinetic term has a physical mass of the form
\be\label{rshier}
m_{phys} = m_0 e^{-\sigma(z_i)} \, ,
\ee
where $z_i$ denotes the location of our brane world along the bulk dimension. If $m_0$
has a (natural) magnitude around the (reduced) four-dimensional Planck mass $2 \times 10^{18}$~GeV, 
(\ref{rshier}) may generate a large hierarchy between the Planck scale
and the particle masses $m_{phys}$ in our world, 
depending on the size of the exponent $\sigma(z_i)$ in the warp factor.
In an RS scenario containing just two branes with opposite tensions, 
our world is identified with the negative tension brane located at $z_i = r_c \pi$ with $\sigma = - k |z|, k > 0$,
and the desired hierarchy is obtained for $k r_c \pi \simeq \mathcal{O}\left(50\right)$.
The exponent $\sigma(z_i)$ is positive in the RS model, so the exponential factor can
only decrease the mass with respect to $m_0$.

A more complex scenario was proposed in~\cite{MR}, involving many brane worlds
and higher-order curvature terms of Gauss-Bonnet type in the bulk:
\bea
S= S_5 + S_4 \, ,
\label{s5s4}
\eea
where $S_5$:
\bea
 S_5 &=&
\int d^5x \sqrt{-g} \left[ -R -\frac{4}{3}\left(\nabla_\mu \Phi \right)^2
+ f(\Phi) \left(R^2 -4 R_{\mu\nu}^2 +
R_{\mu\nu\rho\sigma}^2\right) \right.\nonumber \\
&~& + \xi(z) e^{\zeta \Phi}
+ \left. c_2~f(\Phi)\left(\nabla _\mu \Phi \right)^4 + \dots \right] \, ,
\label{actionGB}
\eea
where $\Phi (z)$ is the dilaton field and the $\dots$ denote other contractions
of the four-derivative dilaton terms that are not relevant, since they can be removed
by appropriate field redefinitions that leave perturbative string amplitudes invariant.
In this case there the bulk Einstein equations have
exact solutions in which our world can be identified with a 
positive-tension brane and, moreover, for the exponent in the warp factor to take \emph{negative} values, $\sigma(z_i) < 0$, 
which is more important for our purposes as it introduces an inverse RS hierarchy.

The action (\ref{s5s4}, \ref{actionGB}) is compatible with closed-string amplitude computations
in five-dimensional space-times, as needed because we assume closed-string propagation in the bulk.
In the stringy case one has~\cite{MR}
\begin{equation}
f(\Phi)=\lambda~e^{\theta\Phi}~,~
\lambda =\alpha '/8g_s^2 > 0~,~c_2=\frac{16}{9}\frac{D-4}{D-2}, \;
\zeta=-\theta=\frac{4}{\sqrt{3(D-2)}} \, ,
\end{equation}
where $\alpha ' = 1/M_s^2$ is the Regge slope, 
$M_s$ is the string mass scale, $g_s$ is the string coupling and the number of space-time
dimensions is $D =5$.

The four-dimensional part $S_4$ of the action (\ref{s5s4}) is given by
\bea
 S_4 = \sum_{i} \int d^4x \sqrt{-g_{(4)}} e^{\omega \Phi} v(z_i) \, ,
\label{s4}
\eea
where
\bea
g_{(4)}^{\mu\nu}= \left(
\begin{array}{l}
g^{\mu\nu}\, , \,\mu,\nu<5\\
0 \; \; ,\, \mbox{otherwise}\\
\end{array}\right) \, .
\eea
and the sum over $i$ extends over D-branes located at points $z=z_i$
along the fifth dimension. Embedding the model (\ref{bare}) in such a scenario, we
identify (\ref{bare}) with the effective four-dimensional field-theory Lagrangian that describes the 
low-energy dynamics of open strings (representing photons) with their ends attached to 
our physical world-brane.

\begin{figure}[ht]
\begin{center}
\includegraphics[width=7cm]{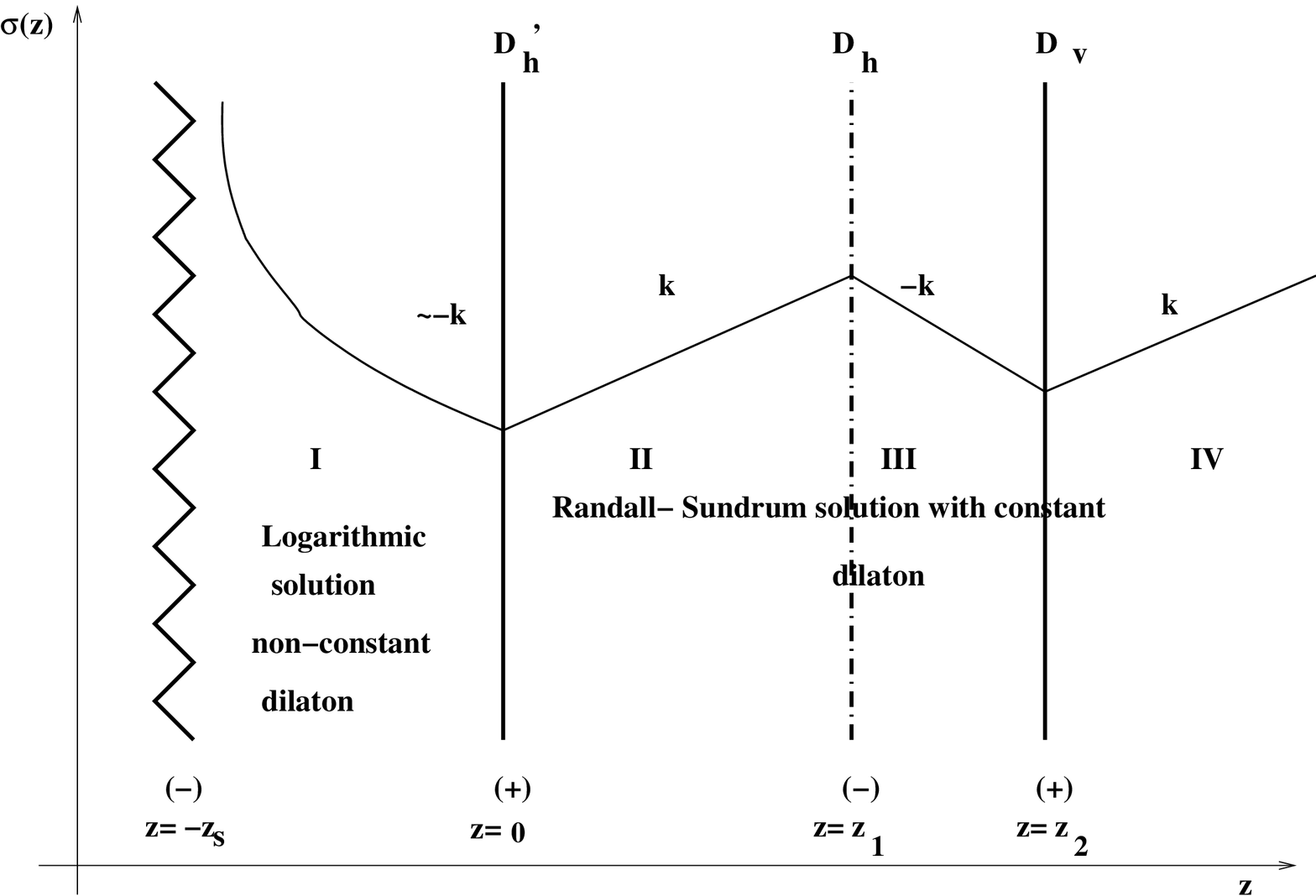} \hfill
\includegraphics[width=7cm]{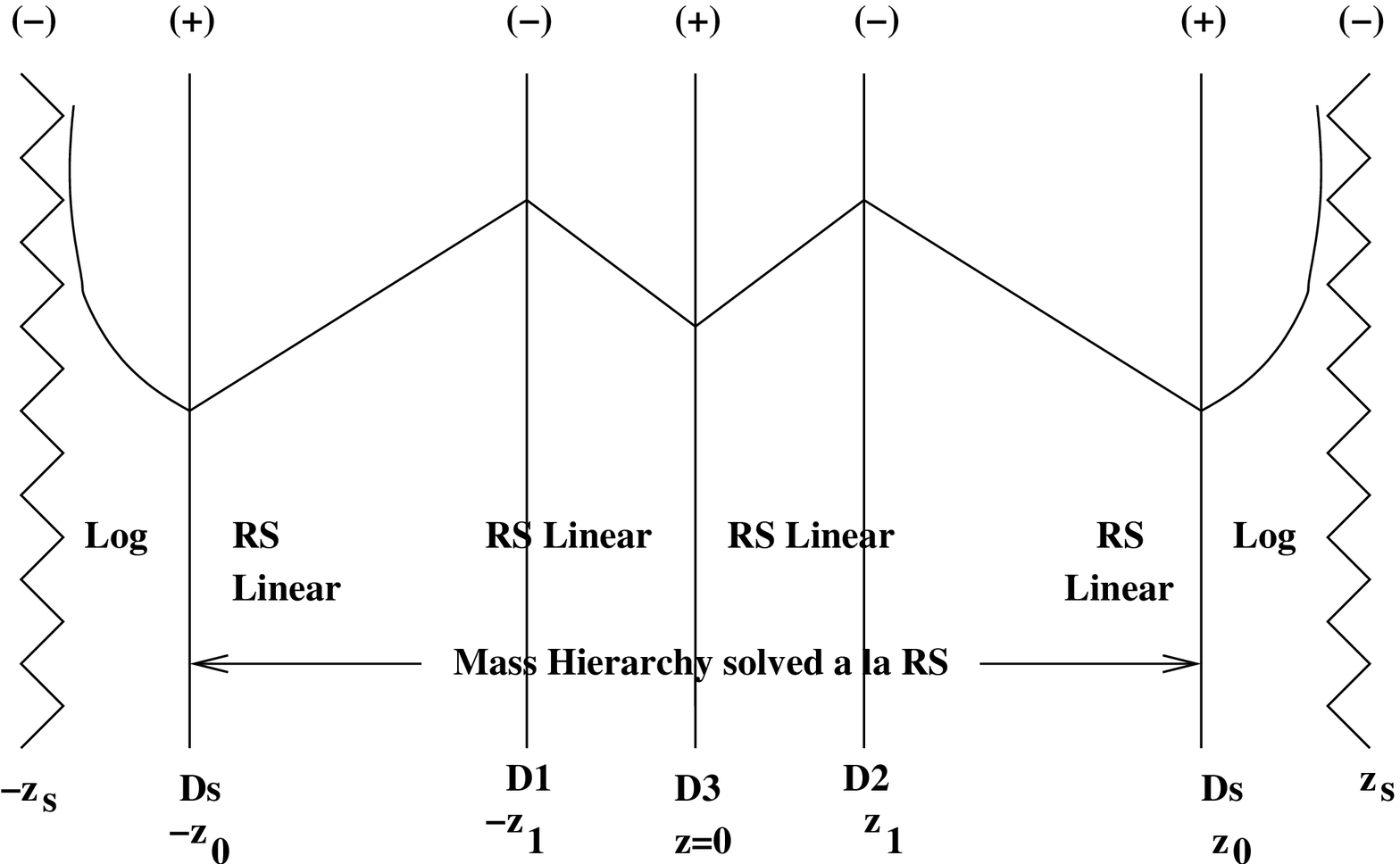}
\caption{\it {Left panel}:
A multi-brane scenario in which our world is represented
by a positive-tension brane at $z=z_2$, accompanied by branes of
alternating-sign tensions on the left, which shield a
bulk naked singularity that may be thought of
as a limiting (singular) case of a negative tension brane.
The bulk dimension extends to infinity to the right of the brane world.
{Right panel}: A multi-brane scenario in which our world is represented
by a positive tension brane at $z=0$, surrounded on both sides by branes
with alternating-sign tensions
that shield two symmetrically-positioned
bulk naked singularities.}
\label{fig:3}
\end{center}
\end{figure}

The analysis of \cite{MR}, to which we refer the interested reader for details, has demonstrated
that assuming a warped five-dimensional geometry of the form (\ref{warp})
there is an exact multi-brane solution for the actions (\ref{s5s4}), (\ref{actionGB}) and (\ref{s4})
that is depicted in Fig.~\ref{fig:3}, which involves bulk singularities restricting dynamically 
the available bulk space. In the bulk regions adjacent to the bulk singularities: $z \sim z_s$
the warp factor has a logarithmic solution: 
\begin{equation}
\sigma (z) \; = \; \sigma_2 + \sigma_1 {\rm log}|z -z_s|~,
\end{equation}
while in the other segments of the bulk space the various brane worlds induce linear solutions
\begin{equation}
\sigma(z) \; = \; \sigma_0 + k z~,
\end{equation}
with the parameter $k$ alternating in sign between adjacent segments of the bulk space, 
as indicated in Fig.~\ref{fig:3}. A consistent scenario is obtained by matching the various solutions on each brane.

The detailed analysis of \cite{MR} derived solutions of the low-energy 
gravitational bulk equations with mass hierarchies of the form:
\be\label{hierarchy2}
       m_{phys} = m_0\,e^{k(2z_1-z_2)}~, \qquad k=\sqrt{\frac{2}{3}}\, g_s M_s > 0~,
\ee
where $z_2$ is the location of our world ($z_2=0$ in the symmetric scenario shown in the 
right panel of Fig.~\ref{fig:3}). The bulk string scale, $M_s$, is in general an arbitrary scale in 
string theory that may be very different from the four-dimensional Planck scale $M_P$.
In this scenario, however, $M_P$ may not be very different from $M_s$, {e.g.},
it can be of order~\cite{MR} $M_P \sim M_s/\sqrt{g_s}$. We stress again that,
in these scenarios, our physical world is a \emph{positive}-tension brane.

Clearly it is possible in such a set-up to have {inverse RS hierarchies}, 
by arranging appropriately the positions of the various branes. For instance, we may have $z_2 < 2z_1 $.
Identifying $m_0$ in (\ref{hierarchy2}) with our dynamically-generated gauge-invariant mass
(\ref{dynmass2}), the physical mass in our brane world would be:
\be\label{finalmass}
m_{dyn} = M {\rm exp}\left(-\frac{2\pi}{3\alpha_V} + \sqrt{\frac{2}{3}}\,g_s M_s |z_2 - 2z_1|\right) \, .
\ee
Singlet fermion masses of the desired phenomenological magnitude can be obtained by
arranging appropriately the distance $|z_2 - 2z_1|$ in the arrangements shown in Fig.~\ref{fig:3}~\footnote{In
the case of the symmetric arrangement in the right panel of Fig.~\ref{fig:3},
our brane world is located at $z_2=0$.}.

As an indication, if we assume that $g_s^2/4 \pi = 1/20$, a value that is typical in
string phenomenology, then the singlet fermion mass due to foam alone would be of order 
\begin{equation}
\label{number}
m_{dyn} \; = \; M e^{- \frac{2\pi}{3 \alpha_V}} \; = \; 10^{-6.37/n_D} M \, .
\end{equation} 
If $n_D = {\cal O}(1)$ per Planck volume and the mass scale $M$ appearing in (\ref{finalmass}, \ref{number})
is ${\cal O}(10^{19})$~GeV, the value of the singlet fermion mass is
\begin{equation}
\label{final}
m_{dyn} \; = \; {\cal O}(10^{13})~{\rm GeV} \, .
\end{equation}
This is typical of the masses of singlet right-handed neutrinos that are postulated in
seesaw models of light neutrino masses~\footnote{It is also in the range postulated for the
inflaton in a wide range of models of cosmological inflation, so in a supersymmetric
extension of our scenario the inflaton could also be identified with the superpartner of the singlet
fermion.}.

\section{Conclusions and Outlook \label{sec:concl} }

We have discussed in this paper the possible role of D-doam in generating dynamically
masses for singlet fermions. An obvious application is to right-handed neutrinos that 
then - via a seesaw mechanism - are responsible for generating the small masses of the active neutrinos
in the Standard Model. As discussed at the end of the previous Section, this
possibility is well within the range of uncertainty in the dynamical
singlet fermion mass. We note that in this approach Lorentz invariance is violated at the foamy level, but this Lorentz
violation is hidden in the effective low-energy field theory, so the phenomenology of neutrino physics
at accessible energies resembles that in a conventional seesaw model.

There is considerable flexibility in the magnitude of the singlet fermion mass, which is suppressed
hierarchically by the coupling $\alpha_V$ whose value is poorly constrained, but may be
enhanced hierarchically by geometrical effects in a multi-brane-world scenario, as discussed
in the previous Section. As commented above, we leave to future work the estimation of $\alpha_V$ in
specific models.

Before closing we recall that the r\^ole of D-foam in the dynamical generation of charged fermions 
was discussed in \cite{mavrolorentz} in a spirit similar to that presented here.
However, there are some important differences.
In the charged-fermion case, the r\^ole of the coupling ${\tilde g}_V$ was played by the 
electrical charge of the fermion, which implied that, for realistic cases, the foam-induced 
dynamical mass was very much suppressed. However, the dynamical vector field $\mathcal{A}_\mu$
discussed here in the singlet neutral fermion case does not couple to a charged fermion, because of charge conservation,
and there is no such suppression.

As discussed in the previous Section, in addition to the coupling $\alpha_V$ to the dynamical
vector field $\mathcal{A}_\mu$, the singlet fermion mass may be enhanced geometrically in a
multi-brane-world scenario. This is a possibility also in the charged-fermion case, but
realistic masses  could be obtained for only one species in that case:
the diversity of the masses of the quarks and leptons of the Standard Model could
not be obtained via the foam mechanism. On the other hand, this possibility is open
in the case of singlet fermions, via appropriate choices of the couplings ${\tilde g}_V$
and the geometrical factors, which are species dependent.

\section*{Acknowledgements}

The work of JE and NEM is partially supported by the
U.K. Science and Technology Facilities Council (STFC) via the Grant ST/L000326/1,  
while that of DVN is partially supported by DOE grant DE-FG02-13ER42020 and
in part by the Alexander S. Onassis Public Benefit Foundation.


\newpage
\begin{thebibliography}{99}

\bibitem{wheeler} J.~A.~Wheeler,
  Phys.\ Rev.\  {\bf 97}, 511 (1955).
  doi:10.1103/PhysRev.97.511


\bibitem{nature} G.~Amelino-Camelia, J.~R.~Ellis, N.~E.~Mavromatos and D.~V.~Nanopoulos,
  Int.\ J.\ Mod.\ Phys.\ A {\bf 12}, 607 (1997)
  doi:10.1142/S0217751X97000566
  [hep-th/9605211];
G.~Amelino-Camelia, J.~R.~Ellis, N.~E.~Mavromatos, D.~V.~Nanopoulos and S.~Sarkar,
  Nature {\bf 393}, 763 (1998)
  doi:10.1038/31647
  [astro-ph/9712103].

\bibitem{notQFT}
S.~W.~Hawking,
  Commun.\ Math.\ Phys.\  {\bf 87} (1982) 395
  doi:10.1007/BF01206031;
  J.~R.~Ellis, J.~S.~Hagelin, D.~V.~Nanopoulos and M.~Srednicki,
  Nucl.\ Phys.\ B {\bf 241} (1984) 381.
  doi:10.1016/0550-3213(84)90053-1

\bibitem{kostel} D.~Colladay and V.~A.~Kostelecky,
  Phys.\ Rev.\  D {\bf 55}, 6760 (1997)
  [arXiv:hep-ph/9703464];
V.~A.~Kostelecky and S.~Samuel,
  Phys.\ Rev.\  D {\bf 40}, 1886 (1989);
 V.~A.~Kostelecky,
  arXiv:0802.0581 [gr-qc], and references therein.
  J.~Bernab\'eu, F.~J.~Botella and M.~Nebot,
  JHEP {\bf 1606}, 100 (2016)
  doi:10.1007/JHEP06(2016)100
  [arXiv:1605.03925 [hep-ph]].

\bibitem{sarkarbeny} J.~Bernab\'eu, N.~E.~Mavromatos, J.~Papavassiliou,
  Phys.\ Rev.\ Lett.\  {\bf 92}, 131601 (2004).
  [hep-ph/0310180];
  E.~Alvarez, J.~Bernab\'eu and M.~Nebot,
  JHEP {\bf 0611}, 087 (2006)
  doi:10.1088/1126-6708/2006/11/087
  [hep-ph/0605211];
J.~Bernab\'eu, N.~E.~Mavromatos, S.~Sarkar,
  Phys.\ Rev.\  {\bf D74}, 045014 (2006).
  [hep-th/0606137];
 J.~Bernab\'eu, F.~J.~Botella, N.~E.~Mavromatos and M.~Nebot,
  arXiv:1612.05652 [hep-ph];
  N.~E.~Mavromatos,
  Found.\ Phys.\  {\bf 40}, 917-960 (2010).
  [arXiv:0906.2712 [hep-th]] and references therein.


\bibitem{LVbounds} V.~A.~Kostelecky and N.~Russell,
  Rev.\ Mod.\ Phys.\  {\bf 83}, 11 (2011)
  doi:10.1103/RevModPhys.83.11
  [arXiv:0801.0287 [hep-ph]].




\bibitem{grav} 
  J.~Ellis, N.~E.~Mavromatos and D.~V.~Nanopoulos,
  Mod.\ Phys.\ Lett.\ A {\bf 31}, no. 26, 1675001 (2016)
  doi:10.1142/S0217732316750018
  [arXiv:1602.04764 [gr-qc]].



\bibitem{dfoam} J.~R.~Ellis, N.~E.~Mavromatos and D.~V.~Nanopoulos,
 Gen.\ Rel.\ Grav.\ \textbf{32}, 127 (2000); 
 Phys.\ Rev.\ D \textbf{61}, 027503 (2000); 
 Phys.\ Rev.\ D \textbf{62}, 084019 (2000).

\bibitem{westmuckett}  J.~R.~Ellis, N.~E.~Mavromatos
and M.~Westmuckett, 
Phys.\ Rev.\ D \textbf{70}, 044036 (2004); 
\emph{ibid.} \textbf{71}, 106006 (2005).

\bibitem{sakharov} J.~R.~Ellis, N.~E.~Mavromatos and A.~S.~Sakharov,
  Astropart.\ Phys.\  {\bf 20}, 669 (2004)
  doi:10.1016/j.astropartphys.2003.12.001
  [astro-ph/0308403];
 J.~R.~Ellis, N.~E.~Mavromatos, D.~V.~Nanopoulos and A.~S.~Sakharov,
  Nature {\bf 428}, 386 (2004)
  doi:10.1038/nature02481
  [astro-ph/0309144];
  Int.\ J.\ Mod.\ Phys.\ A {\bf 19}, 4413 (2004)
  doi:10.1142/S0217751X04019780
  [gr-qc/0312044].


\bibitem{emnnewuncert} J.~R.~Ellis, N.~E.~Mavromatos and D.~V.~Nanopoulos,
 Phys.\ Lett.\ B \textbf{665}, 412 (2008); 
  Int.\ J.\ Mod.\ Phys.\ A {\bf 26}, 2243 (2011)
  doi:10.1142/S0217751X11053353
  [arXiv:0912.3428 [astro-ph.CO]].
  Phys.\ Lett.\  B {\bf 694}, 61 (2010)
  [arXiv:1004.4167 [astro-ph.HE]].


\bibitem{li} T.~Li, N.~E.~Mavromatos, D.~V.~Nanopoulos and D.~Xie,
 Phys.\ Lett.\ B \textbf{679}, 407 (2009). 

\bibitem{string} J.~Polchinski,
  String theory. Vols.  1 \& 2 (Cambridge University Press (2007-12-19))
   SBN: 9780511252273 (eBook), 9780521672276 (Print), 9780521633031 (Print)   
    ISBN: 9780511252280 (eBook), 9780521633048 (Print), 9780521672283 (Print).





\bibitem{mavrolorentz} N.~E.~Mavromatos,
  Phys.\ Rev.\ D {\bf 83}, 025018 (2011)
  doi:10.1103/PhysRevD.83.025018
  [arXiv:1011.3528 [hep-ph]].
  
 
  
  
 \bibitem{alexandre} J.~Alexandre,
  arXiv:1009.5834 [hep-ph];
 J.~Alexandre and A.~Vergou,
  Phys.\ Rev.\ D {\bf 83}, 125008 (2011)
  doi:10.1103/PhysRevD.83.125008
  [arXiv:1103.2701 [hep-th]].


\bibitem{leite} J.~Alexandre, J.~Leite and N.~E.~Mavromatos,
  Phys.\ Rev.\ D {\bf 87}, no. 12, 125029 (2013)
  doi:10.1103/PhysRevD.87.125029
  [arXiv:1304.7706 [hep-ph]].


\bibitem{szabo}  N.~E.~Mavromatos and R.~J.~Szabo, 
 Phys.\ Rev.\ D \textbf{59}, 104018 (1999) {[}arXiv:hep-th/9808124{]};
see also J.~R.~Ellis, N.~E.~Mavromatos and D.~V.~Nanopoulos,
  Mod.\ Phys.\ Lett.\  A {\bf 10}, 1685 (1995)
  [arXiv:hep-th/9503162];
G.~Amelino-Camelia, J.~R.~Ellis, N.~E.~Mavromatos and D.~V.~Nanopoulos,
  Mod.\ Phys.\ Lett.\  A {\bf 12}, 2029 (1997)
  [arXiv:hep-th/9701144].


\bibitem{finsler} See, for instance D. Bao, S.~S.~Chern and Z.~Shen, \emph{An introduction
to Finsler Geometry} (Springer-Verlag, NY, 2000).
In the context of D-particle foam, such Finsler-type metrics were first derived in
J.~R.~Ellis, N.~E.~Mavromatos and D.~V.~Nanopoulos,
  Int.\ J.\ Mod.\ Phys.\  A {\bf 13}, 1059 (1998)
  [arXiv:hep-th/9609238].
For a short review of this topic, see N.~E.~Mavromatos,
  PoS {\bf QG-PH}, 027 (2007)
  [arXiv:0708.2250 [hep-th]] and references therein.
Finsler metrics have been previously suggested in string theory, but in a different context, see 
S.~I.~Vacaru,
  arXiv:hep-th/0211068 and
  arXiv:hep-th/0310132.
In a field-theoretical context, such metrics have been discussed, among other works, in
 G.~Y.~Bogoslovsky,
  arXiv:0706.2621 [gr-qc];
  SIGMA {\bf 4}, 045 (2008)
  doi:10.3842/SIGMA.2008.045
  [arXiv:0712.1718 [hep-th]];
G.~W.~Gibbons, J.~Gomis and C.~N.~Pope,
  Phys.\ Rev.\  D {\bf 76}, 081701 (2007)
  [arXiv:0707.2174 [hep-th]];
  A.~P.~Kouretsis, M.~Stathakopoulos and P.~C.~Stavrinos,
  Phys.\ Rev.\ D {\bf 79}, 104011 (2009)
  doi:10.1103/PhysRevD.79.104011
  [arXiv:0810.3267 [gr-qc]],
  Phys.\ Rev.\ D {\bf 82}, 064035 (2010)
  doi:10.1103/PhysRevD.82.064035
  [arXiv:1003.5640 [gr-qc]] and
  Math.\ Methods Appl.\ Sci.\  {\bf 37}, 223 (2014)
  doi:10.1002/mma.2919
  [arXiv:1301.7652 [gr-qc]];
L.~Sindoni,
  Phys.\ Rev.\  D {\bf 77}, 124009 (2008)
  [arXiv:0712.3518 [gr-qc]];
M.~Anastasiei and S.~I.~Vacaru,
  J.\ Math.\ Phys.\  {\bf 50}, 013510 (2009)
  [arXiv:0710.3079 [math-ph]].
For discussions in the context of generic phenomenological models of non-standard dispersion relations in quantum gravity and
Lorentz-violating Horava-Lifshitz theories, see
F.~Girelli, S.~Liberati and L.~Sindoni,
  Phys.\ Rev.\  D {\bf 75}, 064015 (2007)
  [arXiv:gr-qc/0611024];
J.~Magueijo and L.~Smolin,
  Class.\ Quant.\ Grav.\  {\bf 21}, 1725 (2004)
  [arXiv:gr-qc/0305055];
J.~Skakala and M.~Visser,
  J.\ Phys.\ Conf.\ Ser.\  {\bf 189}, 012037 (2009)
  doi:10.1088/1742-6596/189/1/012037
  [arXiv:0810.4376 [gr-qc]];
S.~I.~Vacaru,
  Gen.\ Rel.\ Grav.\  {\bf 44}, 1015 (2012)
  doi:10.1007/s10714-011-1324-1
  [arXiv:1010.5457 [math-ph]].


 \bibitem{horava} P.~Horava,
  Phys.\ Rev.\  D {\bf 79}, 084008 (2009)
  [arXiv:0901.3775 [hep-th]];
 T.~P.~Sotiriou, M.~Visser and S.~Weinfurtner,
  Phys.\ Rev.\ Lett.\  {\bf 102}, 251601 (2009)
  [arXiv:0904.4464 [hep-th]];
M.~Visser,
  Phys.\ Rev.\  D {\bf 80}, 025011 (2009)
  [arXiv:0902.0590 [hep-th]] and references therein.

\bibitem{review} For a review on this topi, see N.~E.~Mavromatos,
  Int.\ J.\ Mod.\ Phys.\ A {\bf 25}, 5409 (2010)
  doi:10.1142/S0217751X10050792
  [arXiv:1010.5354 [hep-th]], and references therein.

\bibitem{vergou} N.~E.~Mavromatos, S.~Sarkar and A.~Vergou,
  Phys.\ Lett.\ B {\bf 696}, 300 (2011)
  doi:10.1016/j.physletb.2010.12.045
  [arXiv:1009.2880 [hep-th]];
N.~E.~Mavromatos, V.~A.~Mitsou, S.~Sarkar and A.~Vergou,
  Eur.\ Phys.\ J.\ C {\bf 72}, 1956 (2012)
  doi:10.1140/epjc/s10052-012-1956-7
  [arXiv:1012.4094 [hep-ph]];
 N.~E.~Mavromatos, M.~Sakellariadou and M.~F.~Yusaf,
  JCAP {\bf 1303}, 015 (2013)
  doi:10.1088/1475-7516/2013/03/015
  [arXiv:1211.1726 [hep-th]];
 T.~Elghozi, N.~E.~Mavromatos, M.~Sakellariadou and M.~F.~Yusaf,
  JCAP {\bf 1602}, no. 02, 060 (2016)
  doi:10.1088/1475-7516/2016/02/060
  [arXiv:1512.03331 [hep-th]].



\bibitem{arkani} N.~Arkani-Hamed, S.~Dimopoulos, G.~R.~Dvali and J.~March-Russell,
  Phys.\ Rev.\ D {\bf 65}, 024032 (2001)
  doi:10.1103/PhysRevD.65.024032
  [hep-ph/9811448].


\bibitem{rizos} I.~Antoniadis, E.~Kiritsis, J.~Rizos and T.~N.~Tomaras,
  Nucl.\ Phys.\ B {\bf 660}, 81 (2003)
  doi:10.1016/S0550-3213(03)00256-6
  [hep-th/0210263].



\bibitem{recoil} I.~I.~Kogan, N.~E.~Mavromatos and J.~F.~Wheater,
  Phys.\ Lett.\  B {\bf 387}, 483 (1996);
J.~R.~Ellis, N.~E.~Mavromatos and D.~V.~Nanopoulos,
  Int.\ J.\ Mod.\ Phys.\  A {\bf 13}, 1059 (1998).


\bibitem{lcft} For a very partial list of works relevant to our discussion here, see V.~Gurarie,
  Nucl.\ Phys.\  B {\bf 410}, 535 (1993)
  [arXiv:hep-th/9303160];
J.~S.~Caux, I.~I.~Kogan and A.~M.~Tsvelik,
  Nucl.\ Phys.\  B {\bf 466}, 444 (1996)
  [arXiv:hep-th/9511134];
I.~I.~Kogan and N.~E.~Mavromatos,
  Phys.\ Lett.\  B {\bf 375}, 111 (1996)
  [arXiv:hep-th/9512210];
 M.~A.~I.~Flohr,
  arXiv:hep-th/0407003;
J.~Fuchs, S.~Hwang, A.~M.~Semikhatov and I.~Y.~Tipunin,
  Commun.\ Math.\ Phys.\  {\bf 247}, 713 (2004)
  [arXiv:hep-th/0306274];
M.~S.~Movahed, M.~Saadat and M.~Reza Rahimi Tabar,
  Nucl.\ Phys.\  B {\bf 707}, 405 (2005)
  [arXiv:cond-mat/0409486], and references therein.


\bibitem{sarkar}  N.E.~Mavromatos and Sarben Sarkar,
 Physical\ Review\ D \textbf{72}, 065016 (2005) {[}arXiv:hep-th/0506242{]}.

\bibitem{tomboulis} For a partial list of references see: J.~D.~Bjorken,
  Annals Phys.\  {\bf 24}, 174 (1963).
  doi:10.1016/0003-4916(63)90069-1
 G.~S.~Guralnik,
  Phys.\ Rev.\  {\bf 136}, B1404 (1964).
  doi:10.1103/PhysRev.136.B1404
  R.~N.~Sen and C.~Weil,
  Nuovo Cim.\ A {\bf 6}, 581 (1971).
  doi:10.1007/BF02723389
  A.~Kovner and B.~Rosenstein,
  Phys.\ Rev.\ D {\bf 49}, 5571 (1994)
  doi:10.1103/PhysRevD.49.5571
  [hep-th/9210154].
P.~Kraus and E.~T.~Tomboulis,
  Phys.\ Rev.\ D {\bf 66}, 045015 (2002)
  doi:10.1103/PhysRevD.66.045015
  [hep-th/0203221].
  J.~L.~Chkareuli, C.~D.~Froggatt, R.~N.~Mohapatra and H.~B.~Nielsen,
  hep-th/0412225.
C.~A.~Escobar and L.~F.~Urrutia,
  Phys.\ Rev.\ D {\bf 92}, no. 2, 025042 (2015)
  doi:10.1103/PhysRevD.92.025042
  [arXiv:1507.07801 [hep-ph]].

\bibitem{seibergwitten} N.~Seiberg and E.~Witten,
  JHEP {\bf 9909}, 032 (1999)
  [arXiv:hep-th/9908142].



\bibitem{sussk1} N.~Seiberg, L.~Susskind and N.~Toumbas,
  JHEP {\bf 0006}, 044 (2000).


\bibitem{tseytlin}  See, for instance, A.~A.~Tseytlin,
  arXiv:hep-th/9908105.



\bibitem{szabosusy} N.~E.~Mavromatos, R.~J.~Szabo,
  JHEP {\bf 0301}, 041 (2003).
  [hep-th/0207273];
  JHEP {\bf 0110}, 027 (2001).
  [hep-th/0106259].



\bibitem{gms} V.~P.~Gusynin, V.~A.~Miransky and I.~A.~Shovkovy,
  Found.\ Phys.\  {\bf 30}, 349 (2000) and references therein.


\bibitem{lpole} See, for example, for a lattice attempt to address this issue: M.~Gockeler, R.~Horsley, V.~Linke {\it et al.},
  Phys.\ Rev.\ Lett.\  {\bf 80}, 4119-4122 (1998).
  [hep-th/9712244].


\bibitem{B} V.~P.~Gusynin, V.~A.~Miransky and I.~A.~Shovkovy,
  Phys.\ Rev.\  D {\bf 52} (1995) 4747
  [arXiv:hep-ph/9501304].



\bibitem{miransky}
V.~A.~Miransky,
{\it ``Dynamical symmetry breaking in quantum field theories''},
(World Scientific, Singapore, 1993).


\bibitem{shiu} G.~Shiu and L.~T.~Wang,
  Phys.\ Rev.\ D {\bf 69}, 126007 (2004)
  doi:10.1103/PhysRevD.69.126007
  [hep-ph/0311228].

\bibitem{nuMSM} M.~Shaposhnikov,
  JHEP {\bf 0808} (2008) 008
  [arXiv:0804.4542 [hep-ph]];
  J.\ Phys.\ Conf.\ Ser.\  {\bf 39} (2006) 176 and references therein.



\bibitem{RS} L.~Randall and R.~Sundrum,
  Phys.\ Rev.\ Lett.\  {\bf 83}, 3370 (1999)
  [arXiv:hep-ph/9905221].

\bibitem{MR} N.~E.~Mavromatos and J.~Rizos,
  Int.\ J.\ Mod.\ Phys.\  A {\bf 18}, 57 (2003)
  [arXiv:hep-th/0205299];
  Phys.\ Rev.\  D {\bf 62}, 124004 (2000)
  [arXiv:hep-th/0008074].

\end{thebibliography}
\end{document}